\documentclass[useAMS,usenatbib]{mn2e}
\usepackage{graphicx}
\usepackage{natbib} 
\usepackage{amsmath,amssymb}
\usepackage{bbm}
\usepackage{hyperref}

\newcommand{\PP}{{\mathbf P}}
\def\aj{AJ}%
\def\apj{ApJ}%
\def\apjl{ApJ}%
\def\apjs{ApJS}%
\def\aap{A\&A}%
\def\aaps{A\&AS}%
\def\jcap{J. Cosmology Astropart. Phys.}%
\def\mnras{MNRAS}%
\def\prd{Phys.~Rev.~D}%
\def\nat{Nature}%
\title[Filaments for the SDSS]{Detecting filamentary pattern in the cosmic web:\\
a catalogue of filaments for the SDSS}
\author[E. Tempel, R.~S. Stoica, V.~J. Mart\'inez, L.~J. Liivam\"agi, G.~Castellan, and E. Saar, ]{E. Tempel$^{1,2}$\thanks{E-mail:
elmo.tempel@to.ee}, R.~S. Stoica$^{3,4}$, V.~J. Mart\'inez$^{5,6}$, L.~J. Liivam\"agi$^{1,7}$, G.~Castellan$^{3}$, \newauthor and E. Saar$^{1,8}$\\
$^{1}$Tartu Observatory, Observatooriumi~1, 61602 T\~oravere, Estonia\\
$^{2}$National Institute of Chemical Physics and Biophysics, R\"avala pst 10, Tallinn 10143, Estonia\\
$^{3}$Universit\'e Lille 1, Laboratoire Paul Painlev\'e, 59655 Villeneuve d'Ascq Cedex, France\\
$^{4}$Institut de M\'ecanique C\'eleste et Calcul d'Eph\'em\'erides, Observatoire de Paris, 75014 Paris, France\\
$^{5}$Observatori Astron\`omic, Universitat de Val\`encia, C/ Catedr\`atic Jos\'e Beltr\'an 2,  E-46980, Paterna, Spain\\
$^{6}$Departament d'Astronomia \textbf{i} Astrof\'isica, Universitat de Val\`encia, 46100-Burjassot, Val\`encia, Spain\\
$^{7}$Institute of Physics, University of Tartu, 51010 Tartu, Estonia\\
$^{8}$Estonian Academy of Sciences, Kohtu 6, Tallinn 10130, Estonia}

\begin{document}

%\date{Accepted 2013 Month 0.  Received 2013 Month 0; in original form 2013 Month 0}
\date{Accepted 2013 December 18.  Received 2013 December 18; in original form 2013 October 7}

\pagerange{\pageref{firstpage}--\pageref{lastpage}} \pubyear{2013}

\maketitle

\label{firstpage}

%=============================================================================
\begin{abstract}
The main feature of the spatial large-scale galaxy distribution is its intricate network of
galaxy filaments. This network is spanned by the galaxy locations that can be interpreted as
a three-dimensional point distribution. The global properties of the point process can be
measured by different statistical methods, which, however, do not describe directly the
structure elements. The morphology of the large scale structure, on the other hand, is an important property of
the galaxy distribution. Here we apply an object point process with interactions (the Bisous
model) to trace and extract the filamentary network in the presently largest galaxy
redshift survey, the Sloan Digital Sky Survey (SDSS). We search for filaments in the galaxy
distribution that have a radius of about $0.5~h^{-1}\mathrm{Mpc}$. We divide the detected
network into single filaments and present a public catalogue of filaments. We study the
filament length distribution and show that the longest filaments reach the length of
60~$h^{-1}$Mpc. The filaments contain 35--40\% of the total galaxy luminosity and they cover
roughly 5--8\% of the total volume, in good agreement with $N$-body simulations and previous
observational results.

\end{abstract}


%=============================================================================
\section{Introduction}

Large galaxy redshift surveys reveal that the Universe has a salient weblike structure,
called the cosmic web \citep*{Joeveer:78,Bond:96}. Galaxies and matter in the Universe are
arranged into a complex weblike network of dense compact clusters, elongated filaments,
weak two-dimensional sheets, and huge near-empty voids.

The cosmic web is one of the most intriguing and striking patterns found in nature,
rendering its analysis and characterisation far from trivial. The absence of objective and
quantitative procedures for identifying and isolating clusters, filaments, sheets, and
voids in the large-scale matter distribution has been a major obstacle in investigating the
structure and dynamics of the cosmic web. On the other hand, identification and
quantitative description of the details of the cosmic web is important for a broad range of
cosmological issues. It contains information about the structure formation physics, and is
a rich source of information on the global cosmology. The evolution, structure, and
dynamics of the cosmic web depend on the nature of dark matter and dark energy, and on the
properties of the initial density fluctuations generated in the very early Universe. Thus,
these factors must have left their imprint on the web, on its geometry and topology. Thus probes of the large scale structure, such as wide and deep galaxy surveys, enable us to test
current physical and cosmological theories and improve our understanding of
the Universe.

From an observational point of view there is clear evidence that certain observed
properties of galaxies correlate with their environment. For example, the
morphology-density relation stipulates that elliptical galaxies are found preferentially in
crowded environments and spiral galaxies are found in the field
\citep{Einasto:74,Dressler:80}. The same kind of correlation can be found in terms of the
colours and morphology of galaxies \citep{Blanton:05,Tempel:11a}, their star formation
histories, and ages.

Usually, in environmental studies only the local or global density is used, but various
indications argue for a more intricate connection \citep{Lee:08}. While all morphological
types of galaxies correspond to a well-defined range in density, this alone is not
sufficient to differentiate between them: the connection between density and morphology is
more intricate. It is also known that the spin of dark matter haloes is correlated with the
underlying web elements
\citep{Navarro:04a,Brunino:07,Aragon-Calvo:07,Zhang:09,Hahn:10,Codis:12,Libeskind:12,
Libeskind:13, Aragon-Calvo:13a, Trowland:13}. Observations indicate that the rotation axes of
galaxies are aligned with galaxy filaments \citep*{Trujillo:06,Lee:07,Jones:10,Cervantes-Sodi:10,Tempel:12b,Zhang:13,Tempel:13a}. Comparing
the properties of galaxies with the structure of the cosmic web yield valuable information
about the formation and evolution of galaxies.

Galaxy maps are visually dominated by filaments. Filaments are traced by galaxies and
groups and often occupy the regions between massive clusters
\citep*{Pimbblet:04,Murphy:10,Dietrich:12,Jauzac:12}, however, filaments can also be located in voids \citep{Beygu:13,Rieder:13}. The prominent filamentary channels may contain
up to 40\% of the matter in the Universe \citep{Forero-Romero:09,Jasche:09}. Also,
theoretical studies \citep[e.g.][]{Cen:99} have suggested that around half of the warm gas
in the Universe, presumably accounting for the low-redshift missing baryons
\citep*{Fukugita:98,Viel:05}, is hidden in filaments.

Translating the visual impression of the cosmic web into an algorithm that classifies the
local geometry into different environments is not a trivial task, and much work is being
done in this direction. \citet*{Cautun:13} gives a good overview about the various
structure finding algorithms currently available. Among them are the algorithms based on
the gravitational tidal tensor -- the Hessian of the gravitational potential
\citep{Hahn:07,Lee:08,Bond:10a,Bond:10,Forero-Romero:09,Wang:12}, on the velocity field
\citep{Shandarin:11,Hoffman:12,Wang:13}, \emph{skeleton analysis} \citep*{Novikov:06,Sousbie:08}, \emph{watershed segmentation} \citep{Platen:07,Aragon-Calvo:10a}, the tessellations
\citep{Doroshkevich:97,Gonzalez:10,Sousbie:11a,Sousbie:11,Shandarin:12,Aragon-Calvo:07a,Aragon-Calvo:10,
Aragon-Calvo:13a}, Bayesian sampling of the density field \citep{Jasche:09}, minimal spanning tree \citep{Alpasian:13}, and multi-scale
probability mapping \citep{Smith:12}. All these methods are based on different assumptions
and provide different results. Of course, any environment finder should be evaluated by its
merits. A good algorithm should provide a quantitative classification which agrees with the
visual impression and it should be based on a robust and well-defined numerical scheme.

In this work, the detection of filaments is performed using a marked point process with
interactions, called Bisous model \citep*{Stoica:05}. This model approximates the
filamentary network by a random configuration of small segments or cylinders that interact
and connect while building the network. The model was already successfully applied to
observational data and to mock catalogues \citep*{Stoica:07,Stoica:10}. The filaments found
in these papers delineate well the filaments detected by eye and they were evaluated by Monte Carlo statistical tests. This approach has the
advantage that it works directly with the original point process and does not require
smoothing to create a continuous density field. Our method can be applied to relatively
poorly sampled data sets, as the galaxy maps are; it can be applied both to observations
and simulations. Some of the previously mentioned methods can be applied only to
simulations, which makes their use limited.

Here, our marked point process methodology is adapted in order to apply it to the SDSS
data set. Based on the detection obtained using the Bisous model, filament spines are
extracted and a filaments catalogue is built. This compiled data 
can be further used in order to study the properties of filaments and galaxies therein, 
and their relationship with galaxy clusters. Most
of all the previously cited filaments detection methods are based on the calculus of some
gradient, statistics or other measures characterising locally the filament, followed by a
merging, tracking or filtering procedure. The main advantage of using a marked point process
methodology is that it comes freely with a natural way of integration provided by the
probability theory. In this way a simultaneous morphological and statistical characterisation
of the filamentary pattern is allowed. Completing this approach with the spine detection, connects this probabilistic methodology with the richness and the efficiency of the
deterministic techniques already developed.

The paper is organised as follows. In Sections~\ref{sect:data} and \ref{sect:tools} we
describe the data used and the mathematical tools. In Section~\ref{sec:extract_fils} we
define how we build the simulation and extract the spines of filaments. In
Sections~\ref{sect:results} and \ref{sect:discussion} we present and discuss our results.
The description of the catalogue is given in Appendix~\ref{app:1}.

Throughout this paper we assume the WMAP cosmology: the Hubble constant $H_0 =
100\,h\,\mathrm{km\,s^{-1}Mpc^{-1}}$, the matter density $\Omega_\mathrm{m}=0.27$ and the
dark energy density $\Omega_\Lambda=0.73$ \citep{Komatsu:11}.

%=============================================================================
\section{SDSS data}
\label{sect:data}

Our present study is based on the Sloan Digital Sky Survey (SDSS) data release~8
\citep{York:00,Aihara:11}. The galaxy redshifts are typically accurate to $\sim
30~\mathrm{km}\,\mathrm{s}^{-1}$, making it ideal for studies of the large-scale structure.
We use only the main contiguous area of the survey (the Legacy Survey) and the
spectroscopic galaxy sample as compiled in \citet*{Tempel:12}. The lower Petrosian magnitude
limit for this sample is set to $m_r=17.77$, since for fainter galaxies, the spectroscopic
sample is incomplete \citep{Strauss:02}. To exclude the Local Supercluster from the sample,
the lower CMB-corrected distance limit $z=0.009$ was used. The upper limit was set to
$z=0.155$ ($450~h^{-1}\mathrm{Mpc}$), since at larger distances the sample becomes very
diluted. The sample includes 499340 galaxies.

Due to the peculiar velocities of galaxies, which introduce Doppler effects in the redshift
measurement \citep{Jackson:72,Davis:83,Kaiser:87}, the compact structures in redshift-space
are elongated along the line of sight. This is the so-called \emph{Finger-of-God effect}, 
as first introduced by \citet{Tully:78}.
To find the filamentary structure in the SDSS data, we have to suppress first the redshift
distortions for groups. For that we use the Friends-of-Friends (FoF) groups compiled in
\citet{Tempel:12}; the details of the group finding algorithm are explained in
\citet{Tago:08, Tago:10}. We spherize the groups using the rms sizes of galaxy groups in
the sky and their rms radial velocities as described in \citet*{Liivamagi:12} and
\citet{Tempel:12}. The method is similar to that proposed by \citet{Tegmark:04}. Such a
compression will lead to a better estimate of the density field and can help to find the
real filamentary structure. Nevertheless, this compression may suppress some of the
line-of-sight filaments, since the friend-of-friend group finding algorithms cannot
distinguish between groups and exactly line-of-sight filaments. Thus, unique recovery of
the real-space structures is generally not possible. We note that in principle the
redshift-space distortions can be modelled more accurately, introducing a density-dependent
peculiar velocity sampling scheme \citep*{Kitaura:08,Hess:13}. However, we defer this to
future work, since this will affect only a small number of filaments.

To define the filamentary structure, we use Cartesian coordinates based on the SDSS angular
coordinates $\eta$ and $\lambda$, allowing the most efficient placing of the galaxy sample
cone inside a brick: we used the same coordinates to define the superclusters of galaxies
in \citet{Liivamagi:12}. The galaxy coordinates are calculated as follows:
\begin{eqnarray}
    x&=&-d_\mathrm{gal}\sin{\lambda}\nonumber\\ \label{eq:coordin}
    y&=&d_\mathrm{gal}\cos{\lambda}\cos{\eta}\\ 
    z&=&d_\mathrm{gal}\cos{\lambda}\sin{\eta}\nonumber ,
\end{eqnarray}
where $d_\mathrm{gal}$ is the finger-of-god suppressed co-moving distance to a galaxy.

We refer to \citet{Tempel:12} for a more detailed description of the galaxy sample.

%=============================================================================
\section{Mathematical tools}
\label{sect:tools}

In this section we describe the main tools we use to study and extract the filamentary
pattern of the galaxy distribution in the Universe. First of all, a very short and intuitive
definition of marked point processes is given. For a rigorous study of this
subject we recommend as a starting point \citet*{StoyKendMeck95}, \citet{vanLieshout:00},
and \citet{MollWaag04}. Next, our marked point process based methodology is presented. This
methodology includes: the construction of the Bisous model, a simulation algorithm, and an
optimisation procedure. For a detailed mathematical description together with the necessary
convergence proofs of the method, we recommend \citet{Stoica:05,Stoica:07,Stoica:10}.
 
%=============================================================================
\subsection{Marked point processes}
Point processes are random configuration of points. If the points are labelled using a
random mark, we speak about a marked point process. If the marks are the characteristics of
a random geometrical object, we may say that we have an object point process. These
processes were used by~\cite{MartSaar:02} to study the spatial distribution of galaxies.
The observed galaxies were seen as the realisation of a marked point process, as follows.
The centres of the galaxies were the locations in a point process, whereas the different
characteristics of the galaxies (mass, luminosity, etc.) were the marks associated to the
corresponding locations. The marked point processes mathematical framework allowed the
authors to describe the galaxy population, to define statistical descriptors, and to derive
the corresponding estimators \citep*{Martinez:10}.

The simplest marked point process is the Poisson process. In this process the number of
points is chosen according to a Poisson distribution, while the points are spread
independently uniform in the location space where the marked process lives. Then, to each
point a mark is attached independently identically distributed with respect to the marks
distribution. The previous process is called simple, because the independence assumption
involves no interaction between objects. Such interactions can be defined by means of a
probability density with respect to the reference measure given by the unit intensity
marked Poisson point process \citep{vanLieshout:00,MollWaag04,Stoica:05}.  

%=============================================================================
\subsection{Bisous model}
The marked point process we propose for filamentary detection is different from the ones
already used in cosmology. In fact, we do not model the galaxies, but the structure
outlined by the galaxy positions.

Let $K$ be a cosmological sample of finite volume $0<\nu(K)<\infty$, where a finite number
of galaxies \mbox{$\bmath{d}=\{d_1,d_2,\ldots,d_n\}$} are observed. The galaxies positions
are measured in Cartesian coordinates. For SDSS the Eq.~(\ref{eq:coordin}) defines the
coordinates. The feature we are interested in is the filamentary
network outlined by the galaxies positions.

The main hypothesis of our work is that the filamentary network is made of a random
configuration of connected and aligned cylinders, that is the realisation of a marked point
process. This marked point process is named Bisous model and it was specially designed to
generate and analyse random spatial patterns \citep{Stoica:05a,Stoica:07}. We assume that
locally, galaxies may be grouped together inside a rather small cylinder. We also assume
that such small cylinders may combine to form a filament if neighbouring cylinders are
aligned in similar directions. So, the elements of our marked point process are the centres
of the cylinders and their corresponding geometrical shapes. Cylinders are located in the same volume where galaxies are.

\begin{figure}
	\includegraphics[width=84mm]{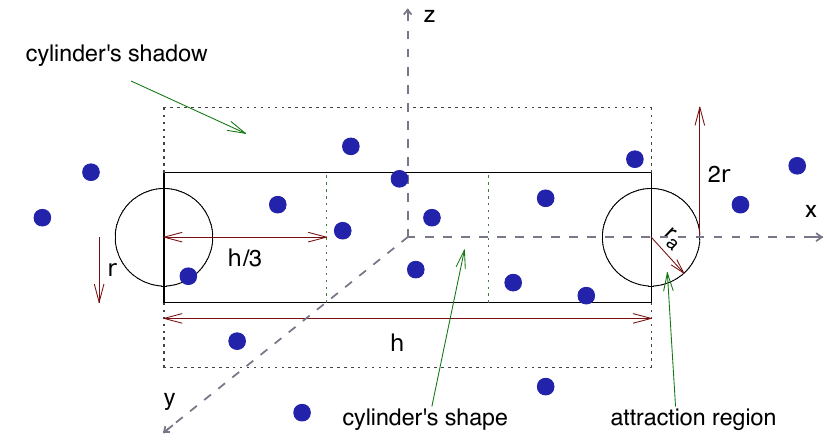}
    \caption{A two-dimensional projection of a cylinder with its shadow within a pattern of 
	galaxies (points). The attraction regions are shown as spheres. The exact shape of the 
	cylinder, its shadow, and attraction regions depend on the model.}
     \label{fig:cylinder}
\end{figure}

A cylinder is an object characterised by its centre $k \in K$ and shape parameters. The
shape parameters of a cylinder are the radius $r$, the length $h$, and the orientation
vector $\omega$. The radius is considered fixed. The length varies uniformly within the interval $[h_{\min}, h_{\max}]$ that will be specified later in this paper. The orientation vector parameters $\omega=\phi(\eta,\tau)$ are uniformly
distributed on $[0,2\pi)\times[0,1]$, such that
\begin{equation}
\omega=\left(\sqrt{1-\tau^2}\cos(\eta),\,\sqrt{1-\tau^2}\sin(\eta),\,\tau\right).
\label{eq:omega}
\end{equation}
Hence, the mark of our process is given by $M=[h_{\min}, h_{\max}] \times [0,2\pi) \times [0,1]$ and its attached uniform distribution $\nu_{M}$. 
We denote the cylinder by $s(y)=s(k,r,h,\omega) \subset K$.

Two extremity rigid points (end points) are attached to each cylinder $s(y)$. Around each
of these points a sphere of the radius $r_\mathrm{a}$ is centred. These two spheres form an
attraction region that are used to define connectivity and alignment rules for cylinders
(see Sect.~\ref{sect:intenergy}). We illustrate the basic cylinder in
Fig.~\ref{fig:cylinder}, where it is centred at the coordinate origin and its symmetry axis
is parallel to the $x$-axis. The coordinates of the end points are
\begin{equation}
	e_u=\left(\frac{h}{2}(-1)^{u+1},0,0\right), \qquad u\in\{1,2\}
\end{equation}
and the orientation vector is $\omega=(1,0,0)$.

Let $\bmath{y}=\{y_1=(k_1,m_1),y_2=(k_2,m_2),\ldots,y_n=(k_n,m_n)\}$ be a configuration of
cylinders, where $m_i$ denotes the mark. The unit intensity independently marked Poisson process constructs a
configuration of cylinders as follows. First, the number $n$ of cylinders is chosen
according to a Poisson law of parameter $\nu(K)$. Then the lengths and the orientation vectors are chosen
independently following $\nu_{M}$. Such a configuration
has only very few connected and aligned cylinders. This effect is just a chance product. In
order to obtain random configurations made of connected and aligned cylinders, a model
defined by a probability density is needed. Such a probability density is specified with
respect to the reference Poisson process and it can be written
\begin{equation}
	p(\bmath{y}|\theta) = \frac{\exp\left[-U(\bmath{y}|\theta)\right]}{Z(\theta)}
	\label{eq:probability}
\end{equation}
where $Z(\theta)$ is the normalising constant, $\theta$ is the vector of the model
parameters, and $U(\bmath{y}|\theta)$ is the energy function of the system (its equivalent
in physics is the total Gibbs energy of a system).

We assumed above that locally, galaxies may be grouped together inside a rather small
cylinder, and such small cylinders may combine to form a filament if neighbouring cylinders
are aligned in similar directions.

Following these two ideas the energy function in~(\ref{eq:probability}) can be specified as:
\begin{equation}
	U(\bmath{y}|\theta) = U_\mathrm{d}(\bmath{y}|\theta) + U_\mathrm{i}(\bmath{y}|\theta),
\end{equation}
where $U_\mathrm{d}(\bmath{y}|\theta)$ is the data energy (see Sect.~\ref{sect:dataenergy})
and $U_\mathrm{i}(\bmath{y}|\theta)$ is the interaction energy (see
Sect.~\ref{sect:intenergy}) associated with the first and second assumptions above,
respectively. In fact, the data energy is related to the position of the cylinders in the
galaxy field, whereas the interaction energy is related to the alignment and connection of
the cylinders constructing the filamentary pattern.

Being in the possession of the model, the parameters have to be chosen. Here, the Bayesian
framework is adopted and the parameters are described by a prior law $p(\theta)$
\citep{Stoica:07a,Stoica:07,Stoica:10}. This allows us to write the joint estimator of
the filamentary pattern and the parameters as
\begin{eqnarray} 
(\widehat{\bmath{y}},\widehat{\theta}) & = & \arg\max_{\Omega \times \Psi} p(\bmath{y},\theta)= \arg\max_{\Omega \times \Psi} p(\bmath{y}|\theta)p(\theta) \nonumber \\
& = &\arg\min_{\Omega \times \Psi}\left\{ \frac{U_{\mathrm{d}}(\bmath{y}|\theta)+U_\mathrm{i}(\bmath{y}|\theta)}{Z(\theta)} + \frac{U_\mathrm{p}(\theta)}{Z_\mathrm{p}(\theta)}\right\}, 
\label{estimator_pattern} 
\end{eqnarray} 
where $p(\theta)=\exp[-U_\mathrm{p}(\theta)]/Z_\mathrm{p}(\theta)$ is the prior law for the model
parameters and $\Psi$ is the model parameters space.

The Bayesian framework was preferred, since we believe, that for the problem at hand, it is
much more natural to give a characterisation of the parameters by a probabilistic law,
instead of a fixed value. Nevertheless, even in this case, some tuning of the model based on
trial and error, is needed. The solution we obtain is not unique. In practise, the shape of
the prior law $p(\theta)$ may influence the solution, making the result to look more random
compared with a result obtained for fixed values of parameters. Therefore, we have derived
tools that are able to average the obtained solution and to state that the obtained results
are really due to the data, and not to a random effect of the presented
methodology \citep{Stoica:07a,Stoica:07,Stoica:10}. Full details concerning the set-up of
the method and the analysis of the results obtained are given later in this paper.

The paper continues with the presentation of the energy terms, the simulation technique,
and an optimisation algorithm.

%=============================================================================
\subsection{Data energy}
\label{sect:dataenergy}

The data energy term is related to the local definition of a galactic filament. This is
still an important open problem. Here, we consider that locally, the galaxies positions
form a filament, if they are situated inside a rather small cylinder, while fulfilling
simultaneously several criteria. The first one is that the galaxies positions should be
spread more or less uniformly a long the main symmetry axis of the cylinder. The second one is
that inside a cylinder there should be more galaxies than outside of it, that is inside
the close-by neighbourhood of the cylinder. And finally, in order to avoid some clustering
effect, the galaxies forming the filaments should be encouraged to get aligned as much as
possible along the main symmetry axis of the cylinder.

Under these circumstances, the data energy of a configuration of cylinders $\bmath{y}$ is
defined as the sum of the energy contributions corresponding to each cylinder:
\begin{equation}
	U_\mathrm{d}(\bmath{y}|\theta) = -\sum\limits_{y\in\bmath{y}} v(y),
	\label{eq:dataenergy}
\end{equation}
where $v(y)$ is the potential function associated with the cylinder $y$. This potential
takes into account the previously mentioned criteria and it depends on $\bmath{d}$ (the
field of galaxies) and the model parameters given by $\theta$.

In order to give a mathematical description of these requirements, an extra cylinder is
attached to each cylinder $y$, with exactly the same parameters as $y$, except for the
radius which equals $2r$. Let $\tilde{s}(y)$ be the shadow of $s(y)$ obtained by the
subtraction of the initial cylinder from the bigger cylinder. The cylinder and its shadow
are shown in Fig.~\ref{fig:cylinder}. Then, we divide each cylinder into three equal
volumes along its main symmetry axis, and denote by $s_1(y)$, $s_2(y)$, and $s_3(y)$ their
corresponding shapes.

Let us assume that locally the number of galaxies inside and around a cylinder, follows a
Poisson distribution.

The first criteria requires the ``local uniform spread'' of the galaxies along the main
symmetry axis of the cylinder. Under the Poissonian assumption, let $\lambda_i,
i=1,\ldots,3$ be the intensity parameters of the corresponding distributions for the shape
regions $s_i(y)$. If the underlying Poissonian process is stationary, ``local uniform spread'' requires all the $\lambda_i$ to be equal. However, filaments are lumpy by nature -- e.g., the filaments in \citet{Pimbblet:04} and the well-known Perseus Chain \citep{Joeveer:78a}. To take this into account, we relax the uniformity assumption by requiring $\lambda_i/\lambda_j$ to be  smaller than chosen threshold.

For any two regions $s_i(y)$ and $s_j(y)$ with $i \neq j$, a statistical test can be done
to compare $\lambda_i$ and $\lambda_j$ \citep{PrzyWile40,KrisThom04}. The test is
\begin{equation}
H_{0,i,j} : \frac{\lambda_{i}}{\lambda_{j}} \leq  \rho_\mathrm{u}
\quad \mathrm{against} \quad
H_{\mathrm{a},i,j} : \frac{\lambda_{i}}{\lambda_{j}} > \rho_\mathrm{u},
\end{equation}
for all pairs of indices $i,j\in\{1,\dots,3 | i\neq j \}$, where $\rho_\mathrm{u}\geq 1$ is a given threshold value.

Now, for a given pair $(i,j)$, the observed number of galaxies in $s_i(y)$ and $s_j(y)$ is
$X_i=m$ and $X_j=n$, respectively. Then, the $p$-value for this test is computed using a
binomial law of parameters $n+m$ and $p(\rho_\mathrm{u}) = \rho_\mathrm{u}/(1+\rho_\mathrm{u})$ as follows
\begin{eqnarray}
	p_\mathrm{u}(s_i(y),s_j(y)) \!\! & = & \!\! \PP(X_i \geq m| X_i + X_j = m+n,p(\rho_\mathrm{u})) \nonumber \\
	&=& \PP(\mathrm{Bin}(n+m,p(\rho_\mathrm{u})) \geq m).
\end{eqnarray}
Six such tests are necessary to verify
the ``local uniform spread'' condition. The obtained score is
\begin{equation}
p_\mathrm{u}(y) = \min \{p_\mathrm{u}(s_i(y),s_j(y)), \!\!\quad\! i,j\!\in\!\{1,\dots,3 | i\neq j \}\}.
\end{equation}
Notice, that this global test is equivalent with verifying $1/\rho_u \leq \lambda_i/\lambda_j \leq \rho_u$ with $i < j $. Hence, it guarantees a minimum density for the galaxies inside a cylinder cell.

The second criteria demands for ``locally high density'' of galaxies inside of a cylinder
comparing to the density of galaxies in the close-by neighbourhood of the cylinder. Under
the Poissonian assumption, the test is
\begin{equation}
H_{0} : \frac{\lambda}{\tilde{\lambda}} \geq \rho_\mathrm{h} 
\quad \mathrm{against} \quad
H_\mathrm{a} : \frac{\lambda}{\tilde{\lambda}} < \rho_\mathrm{h},
\end{equation}
with $\rho_\mathrm{h}$ a given threshold value. It is important to notice that the volumes ratio of
$s(y)$ and $\tilde{s}(y)$ plays an important role in choosing the appropriate value of
$\rho_\mathrm{h}$. For instance, if $\rho_\mathrm{h}=1/3$ this tests if the two processes have the same
intensity, if $\rho_\mathrm{h}=1$ this tests if the intensity inside the cylinder is three times
higher than inside its shadow.

If the observed number of galaxies inside $s(y)$ and $\tilde{s}(y)$ is $X=m$ and
$\tilde{X}=n-m$, respectively, the $p$-value of the test is computed using a binomial
distribution of parameters $n$ and $p(\rho_\mathrm{h}) = \rho_\mathrm{h}/(1+\rho_\mathrm{h})$
\begin{eqnarray}
p_\mathrm{h}(y) & = & \PP(X \leq m| X + \tilde{X} = n,p(\rho_\mathrm{h})) = \\
&=& \PP(\mathrm{Bin}(n,p(\rho_\mathrm{h})) \leq m). \nonumber
\end{eqnarray}

%\begin{equation}
%p_{h}(y)= \PP(X < m| X + \tilde{X} = n,p(\rho_h)) = \PP(\textrm{Bin}(n,p(\rho_h)) < m).
%\end{equation}

To take into account both tests simultaneously, the following score is defined
\begin{equation}
p_\mathrm{hyp}(y)=p_\mathrm{u}(y)\cdot p_\mathrm{h}(y).
\end{equation}

The previous tests are based on counts of points in some pre-defined regions. In order to
take into account the spatial distribution of galaxies in a cylinder we define the
cylinder concentration as
\begin{equation}
\sigma^{2}=\frac{1}{n-2}\sum_{j=1}^n \frac{\delta_j^2}{r^2},
\label{eq:cylinderVariance}
\end{equation}
with $n$ the number of galaxies covered by the cylinder, $\delta_j$ the distance from the
$j$th galaxy inside the cylinder to its main symmetry axis, and $r$ is the cylinder
radius. The weight $1/(n-2)$ is chosen to eliminate some pathological cases with too few
points covered by a cylinder (there must be at least three galaxies inside a cylinder).
Clearly, the concentration $\sigma^2$ has a minimum when the symmetry axis of the cylinder
coincides with the least mean square line passing through the cloud of points, given by
the galaxy positions inside the cylinder.

It is important to notice that the use of this term may induce a local Gaussian
assumption. This may be considered contradictory to the Poisson hypothesis previously
used. Nevertheless, the use of these two strategies is complementary: the first two
requirements impose conditions on the number of galaxies inside a cylinder, while the
third one imposes conditions on the spatial distribution of these galaxies. The term given
by~(\ref{eq:cylinderVariance}) is also a very good indicator of a locally high density and
a better estimator for the filament axis.

The potential function $v(y)$ of the cylinder is built using the previous statistical
tests and criteria. Let us assume that for the cylinder $y$, the $p$-value $p_\mathrm{hyp}$ is
computed as previously, and let $\sigma^2(y)$ be the cylinder concentration. We want
$v(y)$ to be maximum for the ``best'' location of the cylinder in the galaxy field. This
allows the definition of the potential function as
\begin{equation}
v(y) = 
\begin{cases}
- \sigma^{2}(y)+c_\mathrm{hyp}\log \left[p_\mathrm{hyp}(y)\right] & \text{if} \quad n \geq n_\mathrm{min} \\
-\infty & \text{if} \quad n < n_\mathrm{min}
\end{cases}
\label{eq:datapotential}
\end{equation}
with $n$ the number of galaxies covered by the cylinder $s(y)$ and $n_\mathrm{min}$ a
given threshold value. Here, the
formula~\eqref{eq:cylinderVariance} suggests $n_\mathrm{min} \geq 3$. The parameter $c_\mathrm{hyp}\geq
0$ is required to make the two terms comparable: this allows to
use these two strategies safely together.

This gives for the data energy defined by Eq.~(\ref{eq:dataenergy}):
\begin{equation}
U_{\mathrm{d}}(\bmath{y}|\theta)  = -\sum_{y \in \bmath{y}}\left\{ c_\mathrm{hyp}\log \left[p_\mathrm{hyp}(y)\right] -\sigma^{2}(y)\right\},
\end{equation}
and for the data term model
\begin{eqnarray}
	\label{eq:dataModel}
p_{\mathrm{d}}(\bmath{y}|\theta) & \propto & \exp\left[-U_{\mathrm{d}}(\bmath{y}|\theta)\right] \nonumber \\
& \propto & \exp\left[-\sum_{y \in \bmath{y}}\sigma^{2}(y)\right]\prod_{y \in \bmath{y}}[p_\mathrm{hyp}(y)]^{c_\mathrm{hyp}}. 
\end{eqnarray}

The data term model (\ref{eq:dataModel}) is a super-position of inhomogeneous Poisson
point processes with respect to the reference measure. Since the number of galaxies is
finite and since the observed window has a limited volume, the term $\sigma^2$ is always
finite. Therefore, the data term model is locally stable, hence it designs a well defined
model that has an integrable probability density. 

One more point has to be retained
concerning the data term. The use of $p$-values for constructing a potential function is
different of the use of the values for ``purely'' statistical tests. In this last situation
Bonferroni or \v{S}id\'ak corrections are required.

%=============================================================================
\subsection{Interaction energy}
\label{sect:intenergy}

\begin{figure}
	\includegraphics[width=84mm]{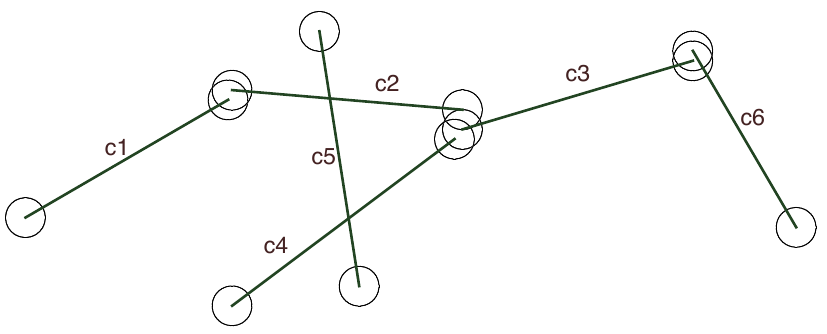}
    \caption{A two dimensional representation of a cylinder configuration: attraction 
	regions are shown with spheres. In this configuration, we observe that the cylinders 
	$c1-c2$, $c2-c3$, and $c3-c4$ are connected. The cylinders $c1$, $c3$, and $c4$ are 
	connected to the network through one end point, while $c2$ is connected to the network 
	through both end points. The cylinders $c5$ and $c6$ are not connected to anything, 
	$c3-c6$ are attracting each other but they are not well aligned, and $c5$ is not 
	attracted to any other cylinder. The cylinder $c5$ is rejecting the cylinders $c2$ and 
	$c4$ (the centres of these cylinders are too close), but as it is rather orthogonal 
	both to $c2$ and $c4$, it is not repulsing them. The cylinders $c2$ and $c4$ reject 
	each other and are not orthogonal, so they form a repulsive pair.}
     \label{fig:cylconf}
\end{figure}

The interaction energy is related to the relative position of the cylinders forming the
network and its expression is as follows
\begin{equation} U_\mathrm{i}(\bmath{y}|\theta)=-n_k(\bmath{y})\log\gamma_k-\sum\limits_{s=0}^2n_s(\bmath{y})\log\gamma_s,
	\label{eq:intenergy}
\end{equation}
where $n_k(\bmath{y})$ is the number of repulsive cylinder pairs and $n_s(\bmath{y})$ is
the number of cylinders connected to the network through $s$ extremity points. The
variables $\log\gamma_k$ and $\log\gamma_s$ are the potentials associated with these
configurations, respectively.

The interaction energy (\ref{eq:intenergy}) is defined in the same way as in
\citet{Stoica:10}.

Two cylinders are considered repulsive, if they are rejecting each other and if they are
not orthogonal. We say that two cylinders $y_1=(k_1,r,h_1,\omega_1)$ and
$y_2=(k_2,r,h_2,\omega_2)$ reject each other if their centres are closer than the minimum
allowed distance between cylinders, $d(k_1,k_2)<0.5(h_1+h_2)-r_\mathrm{a}$. Two cylinders
are considered to be orthogonal if $\lvert\omega_1\cdot\omega_2\rvert\le \tau_\perp$,
where $\cdot$ is the scalar product of the two orientation vectors and
$\tau_\perp\in(0,1)$ is a predefined parameter. So, a certain range of mutual angles is
allowed for cylinders considered to be orthogonal.

Two cylinders are connected if they attract each other, do not reject each other, and are
well aligned. Two cylinders attract each other if the distance between the cylinder end
points is smaller than the interaction radius $r_\mathrm{a}$ (see
Fig.~\ref{fig:cylinder}). Two cylinders are well aligned if
$\lvert\omega_1\cdot\omega_2\rvert\ge 1-\tau_{\|}$, where $\tau_{\|}\in(0,1)$ is a
predefined parameter.

To illustrate these definitions, we show an example configuration of cylinders (in two
dimensions) in Fig.~\ref{fig:cylconf}. Altogether, the configuration at
Fig.~\ref{fig:cylconf} adds to the interaction energy contributions from three
single-connected cylinders ($c1$, $c3$, $c4$), one doubly-connected cylinder ($c2$), two
free cylinders ($c5$, $c6$), and one repulsive cylinder pair ($c2,c4$).

The complete model (\ref{eq:probability}) that includes the definition of the data energy
and of the interaction energy is well defined for parameters
$\gamma_0,\gamma_1,\gamma_2>0$, $c_\mathrm{hyp}\geq 0$, and $\gamma_k\in [0,1]$. The definition of the interactions
and the parameter ranges chosen ensure that the complete model is locally stable
\citep{vanLieshout:00,MollWaag04,Stoica:05}. This property ensures that we can safely use
this model without expecting any dangers (integrability, convergence, numerical stability,
etc). The values of the interaction parameters ($\gamma_s$, $\gamma_k$) and of the data
parameter $c_\mathrm{hyp}$ have to be fixed taking into account the weight of each energy component and
also the underlying galaxy field. If the interaction energy parameters are too strong,
then the filamentary network may appear in location where the galaxies form no filaments.
If the data energy parameters are too strong, then the filamentary pattern will be well
located but not really forming a filamentary network. This is a normal compromise to be
found such as in solution regularisation or Bayesian analysis. The Sect.~\ref{sect:setup}
shows how these parameters were set.

%=============================================================================
\subsection{Simulation of the model and optimisation algorithm}
\label{sect:simulation}

Several Markov chains Monte Carlo (MCMC) techniques are available to simulate marked point
processes: spatial birth-and-death processes, Metropolis-Hastings algorithms, reversible
jump dynamics or more recent exact simulation techniques
\citep{Geyer:94,Green:95,Geyer:99,Kendall:00,vanLieshout:00,vanLieshout:06}.

In this paper, we need to sample from the joint probability density law
$p(\bmath{y},\theta)$. This is done by using an iterative MCMC algorithm. An iteration of
the algorithm consists of two steps. First, a value for the parameter $\theta$ is chosen
with respect to $p(\theta)$. Then, conditionally on $\theta$, a cylinder pattern is
sampled from $p(\bmath{y}|\theta)$ using a Metropolis-Hastings
algorithm~\citep{Geyer:94,Geyer:99}.

The Metropolis-Hastings (MH) algorithm we used is built using three types of moves
\citep{Lieshout:03,Stoica:05,Stoica:07,Stoica:10}. 
\begin{itemize}
	%---------------------
	\item{Birth:} with a probability $p_\mathrm{b}$ a new cylinder $\zeta$, sampled from 
	the birth rate $b(\bmath{y},\zeta)$, is proposed to be added to the present 
	configuration $\bmath{y}$. The new configuration $\bmath{y}\prime = 
	\bmath{y}\cup\{\zeta\}$ is accepted with the probability
	\begin{equation}
		\min \left\{ 1, 
		\frac{p_\mathrm{d}}{p_\mathrm{b}} 
		\frac{d(\bmath{y}\cup\{\zeta\},\zeta)}{b(\bmath{y},\zeta)} 
		\frac{p(\bmath{y}\cup\{\zeta\})}{p(\bmath{y})} 
		\right\}.
		\label{eq:birth_prob}
	\end{equation}
	%---------------------
	\item{Death:} with a probability $p_\mathrm{d}$ a cylinder $\zeta$ from the current 
	configuration $\bmath{y}$ is proposed to be eliminated according to the death proposal 
	$d(\bmath{y},\zeta)$. The role of this move is to ensure the detailed balance of the 
	simulated Markov chain and its convergence towards the equilibrium distribution. The 
	probability of accepting the new configuration $\bmath{y}\prime = 
	\bmath{y}\backslash\{\zeta\}$ is computed reversing the ratio (\ref{eq:birth_prob}).
	%---------------------
	\item{Change:}  with a probability $p_\mathrm{c}$ we randomly choose a cylinder 
	$\zeta_\mathrm{old}$ in the configuration $\bmath{y}$ and propose to slightly change its 
	parameters using uniform proposals. For the selected element, we may change its 
	location within the vicinity $\Delta k$ of its centre and change its orientation within 
	a small angle tolerance $\Delta \omega$ with respect its initial orientation. The new 
	element obtained is $\zeta_\mathrm{new}$. This move improves the mixing properties of the 
	sampling algorithm. The new configuration $\bmath{y}\prime = 
	\bmath{y}\backslash\{\zeta_\mathrm{old}\}\cup\{\zeta_\mathrm{new}\}$ is accepted with the probability
	\begin{equation}
		\min \left\{ 1, 
		\frac{p(\bmath{y}\backslash\{\zeta_\mathrm{old}\}\cup\{\zeta_\mathrm{new}\})}{p(\bmath{y})} 
		\right\}.
		\label{eq:change_prob}
	\end{equation}
\end{itemize}

Some practical details concerning the MH dynamics
implementation are given below. For a complete description we recommend
\citet{Lieshout:03} and \citet{Stoica:05}.

The uniform choices $b(\bmath{y},\zeta)=1/\nu(K)$ and $d(\bmath{y},\zeta)=1/n(\bmath{y})$
are commonly adopted for their simplicity and because they guarantee the necessary
convergence properties of the simulated Markov chain, such as irreducibility, Harris
recurrence, and geometric ergodicity. For the probabilities $p_\mathrm{b}$, $p_\mathrm{d}$,
and $p_\mathrm{c}$, all the convergence properties are preserved as long as
$p_\mathrm{b}+p_\mathrm{d}+p_\mathrm{c} \leq 1$. Here, $\nu(K)$ is the Lebesgue measure
(volume) and $n(\bmath{y})$ is the number of cylinders in the configuration.

Nevertheless, when the model to simulate exhibits complicated interactions, such an update
mechanism built of uniform birth and death proposals may be very slow in practice. Here
the strategy proposed by \citet{Lieshout:03} and \citet{Stoica:05} is adopted. This strategy uses
adapted moves that help the model. In our case, the new cylinder can be added uniformly in
$K$ (the observed volume) or can be randomly connected with the rest of the network. This
mechanism helps to build a connected network and it can be implemented using a non-uniform
mixture for the birth proposal
\begin{equation}
	b(\bmath{y},\zeta)=\frac{p_1}{\nu(K)}+ p_2b_\mathrm{a}(\bmath{y},\zeta),
\end{equation}
with $p_1+p_2=1$ ($p_2$ is the probability to add a connected cylinder) and
$b_\mathrm{a}(\bmath{y},\zeta)$ is a probability density proposing attracting and well
aligned (e.g. connected) cylinders. The expression of $b_\mathrm{a}(\bmath{y},\zeta)$ is
given by
\begin{equation}
	b_\mathrm{a}(\bmath{y},\zeta) = \frac{1}{n[A(\bmath{y})]}\sum_{y\in A(\bmath{y})} \tilde{b}(y,\zeta),
	\label{eq:ba}
\end{equation}
where $A(\bmath{y})$ is the set of cylinders in the configuration $\bmath{y}$ which have
at least one end point able to create connections, and $n[A(\bmath{y})]$ is the number of
such cylinders in the configuration. Note that neglecting the edge effects,
$n[A(\bmath{y})]$ is the number of 0- and 1-connected cylinders in configuration. After
choosing uniformly an object $y$ from the set $A(\bmath{y})$, a new object
$\zeta=(k_\zeta,\omega_\zeta)$ is proposed to be added using the density
\begin{equation}
	\tilde{b}(y,\zeta) = \frac{\mathbbm{1} \{ k\in \tilde{a}(y) \} } 
	{\nu[ \tilde{a}(y)\cap K ]}
	\frac{1}{\tau_\|},
\end{equation}
where $\tilde{a}(y)$ is the region built from the union of attraction balls of $y$ which
are not containing the end of any other attracting cylinder in the configuration
$\bmath{y}$, $\nu[ \tilde{a}(y)\cap K ]$ is the volume of those attraction balls, and
$\mathbbm{1}\{\cdot\}$ is the indicator function that selects the cylinders the new
cylinder $\zeta$ may be connected with. Here, one end point of the proposed connected
cylinder is uniformly distributed in $\tilde{a}(y)$ and the orientation $\omega_\zeta$ is
uniformly chosen to satisfy the well-aligned criterion $\omega_\zeta\cdot\omega_y\geq
1-\tau_\|$. Clearly, the summation in Eq.~(\ref{eq:ba}) is effectively over the cylinders
the new cylinder $\zeta$ can be connected with.

This birth rate leads the model to propose configurations with connected objects much more
often than using the simple uniform proposal. In practice, it is also reasonable to sample
only in the regions where the data potential is defined: $v(y)>-\infty$. Hence, the
Lebesgue measure $\nu(K)$ in this case can be calculated as

%\begin{equation}
%\nu(K) = \int\limits_{k\in K}\mathrm{d}k\, 
%\frac{1}{2\pi}\!\int\limits_{\omega \in M} \mathbbm{1}  
%\left\{v[y(k,\omega)]>-\infty\right\} \mathrm{d}\omega.
%\end{equation}

\begin{equation}
\nu(K) =\!\! \int\limits_{k\in K}
\!\!\mathbbm{1}\!\!\left[\left(
\int\limits_{m \in M}  \!\!\!\!\mathbbm{1}\!  
\left\{v[y(k,m)]\!>\!-\infty\right\}
\mathrm{d}\nu_{M}(m)\right) \!>\! 0
\right]
\!\!\mathrm{d}k
\end{equation}

In order to perform the maximisation of $p(\bmath{y},\theta)$, the previously described
sampling mechanism is integrated into a simulated annealing algorithm. The simulated
annealing is a global optimisation method. It iteratively samples from
$p_n(\bmath{y},\theta) \propto [p(\bmath{y}|\theta)p(\theta)]^{1/T_n}$, while $T_n$ goes
slowly to zero. \citet{Stoica:05} proved the convergence of a simulated annealing
algorithm based on a Metropolis-Hastings dynamics for marked point processes, if a
logarithmic cooling schedule is used. According to this result, the temperature is lowered
as
\begin{equation}
	T_n=\frac{T_0}{\log n+1},
\end{equation}
where $T_0$ is the initial temperature.

%=============================================================================
\section{Extracting and defining the filaments}
\label{sec:extract_fils}

In this section we describe how we set up the experiment and extract the filaments. Our
aim is to use the result obtained using marked point processes to compile a filament
catalogue. Every filament in this catalogue is represented as a spine: a set of points
that define the axis of the filament.

\begin{table}
 \caption{All parameters used to define the model and to extract the filaments. All distances are in $h^{-1}$Mpc.}
 \begin{tabular}{@{}llc}
  \hline\hline
  Param. & Description & Value \\
  \hline
  $r$ & cylinder radius & 0.5 \\
  $h$ & cylinder length & $[3.0,5.0]$ \\
  \hline
  $\rho_\mathrm{h}$ & locally high density & 1.0 \\
  $\rho_\mathrm{u}$ & locally uniform spread & 4.0 \\
  $n_\mathrm{min}$ & minimum number of galaxies & 3 \\
  $c_\mathrm{hyp}$ & hypothesis test coefficient & $[0.7,0.9]$ \\
  \hline
  $r_\mathrm{a}$ & attraction radius & 0.5 \\
  $\tau_\perp$ & orthogonal cylinders & 0.30 \\
  $\tau_{\|}$ & parallel cylinders & 0.15 \\
  $\gamma_k$ & repulsive cylinders & 0 \\
  $\log\gamma_0$ & 0-connected cylinders & $[-2.0,-1.0]$ \\
  $\log\gamma_1$ & 1-connection cylinders & $[-1.0,0.0]$ \\
  $\log\gamma_2$ & 2-connection cylinders & $[0.8,1.8]$ \\
  \hline
  $p_\mathrm{b}/p_\mathrm{d}/p_\mathrm{c}$ & birth/death/change probabilities & 0.5/0.2/0.3 \\
  $p_\mathrm{2}$ & connected birth probability & 0.8 \\
  $\Delta k$ & max shift for change move & 0.2 \\
  $\Delta \omega$ & min cosine for orientation change & 0.95 \\
  $T_0$ & initial temperature & 5.0 \\
  $\delta$ & steps between temperature changes & 100\,000 \\
  $N_\mathrm{iter}$ & number of cycles & 200\,000 \\
  $N_\mathrm{sim}$ & number of simulations & 50 \\
  \hline
  $\mathcal{L}_\mathrm{lim}$ & limiting visit map value for filaments & 0.05 \\
  $\mathcal{D_G}_\mathrm{lim}$ & limiting orientation strength & 0.75 \\
  $\tau_\mathrm{lim}$ & limiting angle for filaments & 0.95 \\
  $\kappa_\mathrm{lim}$ & limiting curvature for filaments & 1.0 \\
  \hline
 \end{tabular}
 \label{tab:params}
\end{table}

%=============================================================================
\subsection{Experimental setup}
\label{sect:setup}

As described above, we use the data set drawn from the SDSS contiguous area. The sample
region $K$ is the observed volume in space.

In order to choose the values for the dimensions of the cylinder, we use the physical
dimensions of galaxy filaments that have been observed in more detail \citep{Pimbblet:04};
we used the same values also in our previous papers \citep{Stoica:07,Stoica:10}: a radius
$r=0.5\,h^{-1}\mathrm{Mpc}$. The same scale has been also used by \citet{Smith:12} and
\citet{Tempel:12b} showed that filaments of this size may influence galaxy evolution, so
this seems to be the most interesting scale for galaxy filaments. Naturally, the nature of
filaments is hierarchical \citep{Shen:06,vandeWeygaert:08a,vandeWeygaert:08,Aragon-Calvo:13} 
and the chosen scale of filaments can be
arbitrary. In the present paper, we aim to detect filaments that have the strongest impact on
galaxy evolution: for that the scale should be relatively small. Taking into account the
data resolution in the SDSS, the scale $r=0.5\,h^{-1}\mathrm{Mpc}$ is the minimal one we
can choose. The length of a cylinder is chosen to be $h=3.0$--$5.0\,h^{-1}\mathrm{Mpc}$,
which is the shortest possible (the ratio of the cylinders length to its diameter is 3:1
to 5:1). The length in this range is considered to be free to more effectively sample the
low number density regions. 

We choose the attraction radius $r_\mathrm{a}=r$, which ensures that the end points of
connected cylinders are not too far apart. For the cosines of the maximum curvature angles
we choose $\tau_\|=0.15$ and $\tau_\perp=0.3$. This allows for a maximum of $\approx
30\degr$ between the direction angles of connected cylinders and considers the cylinders
to be orthogonal, if the angle between their directions is larger than $\approx 70\degr$.

The model parameters $(r,h,r_\mathrm{a})$ influence the detection results. If they are too
low, no filaments will be detected. If they are too high, the detected filaments will be
too wide and/or sparse, and precision will be lost. Still, the precision can be increased
and the influence of model parameters can be minimised, when sets of simulations and visit
maps are used (see Sect.~\ref{sect:spines}): in a certain manner, it will average the
detection results. In this work, the model parameters $(r,h,r_\mathrm{a})$ were chosen
based on a previous knowledge and after a visual inspection of the detected filamentary
pattern.

\begin{figure}
	\includegraphics[width=84mm]{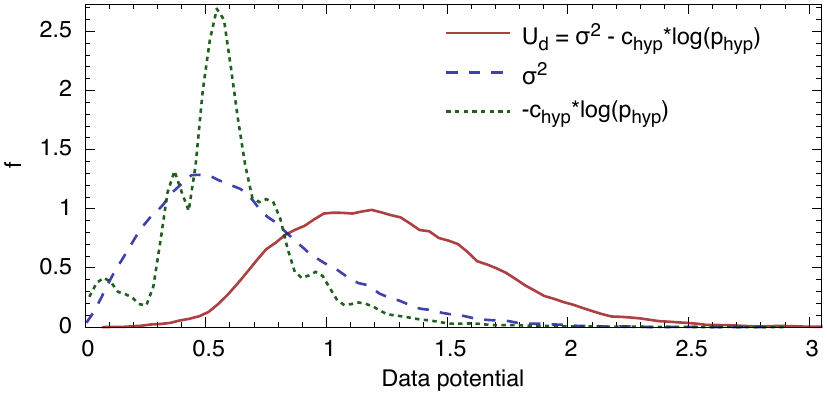}
    \caption{Distribution of cylinder data potentials in a final configuration $\bmath{y}$ 
	(solid red line): the contribution to the data potentials from hypothesis testing 
	(dotted green line) and from concentration (dashed blue line) are also shown. The 
	peaks in the hypothesis testing distribution are caused by a small number of galaxies 
	in a cylinder.}
	\label{fig_dataterm}
\end{figure}

\begin{figure*}
	\includegraphics[width=175mm]{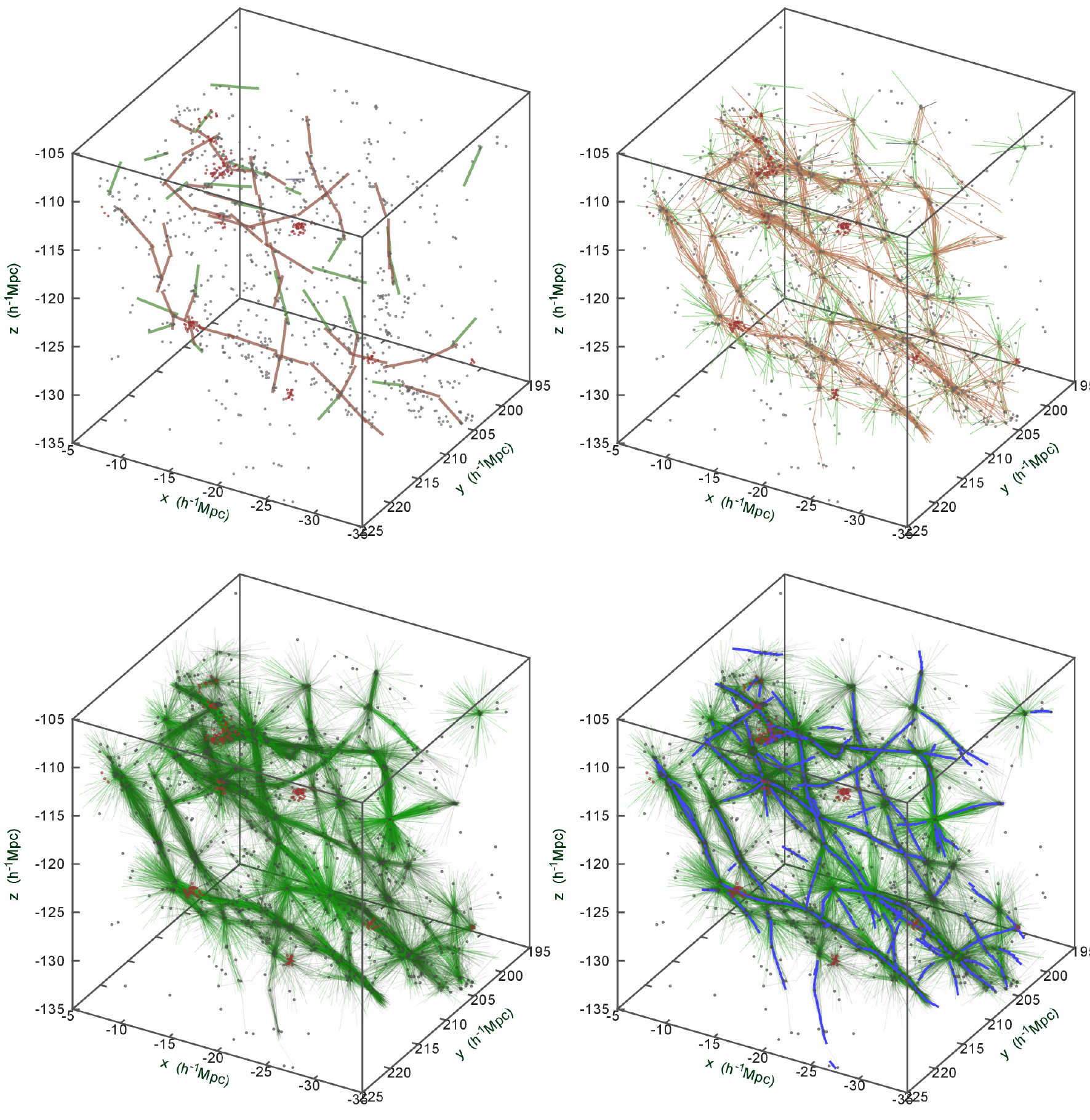}
	\caption{Detected filamentary pattern (cylinder axes) in a small sample volume within 
	a pattern of galaxies (points). \emph{Upper left} panel: single MCMC simulation 
	detecting the filamentary pattern; \emph{upper right} panel: the superposition of $25$ 
	independent simulations (for visual clarity, we show only half of the simulations). 
	Cylinders are colour-coded as following: 2-connected (red), 1-connected (green), and 
	isolated (grey). Galaxies in groups with 10 or more members are shown with red points; 
	other galaxies are shown with grey points. \emph{Lower} panels show the cylinders from 
	1000 realisations (it corresponds to the visit map) used to extract the filament 
	spines; in \emph{lower right} panel, the extracted filament spines are also shown with 
	blue lines. The movie, showing the MCMC in action is available at 
	\url{http://www.aai.ee/~elmo/sdss-filaments/}.}
\label{fig:snapshot}
\end{figure*}

Fixing the data and interaction energy parameters require an ``initial guess'' of the size
of the solution. This guess does not need to be precise. The stochastic algorithmic
``machinery'' will do the job, due to its mathematical theoretical properties. Nevertheless, from a practical point of view, if the
range of parameters allows only very few cylinders in the configuration, then the
detection may be incomplete. On the other hand, if too many cylinders are allowed, then
the detection may contain a lot of false alarms. Attention should be also paid when the
measure units are fixed. A transformation of the measure units induces a transformation of
the model parameters so that the same probabilities are assigned to the same
configurations of objects. Still, a direct relation between the change of the measure
units and the model parameters is not easy to be derived, because of the non-linear
character of the model. Under these circumstances, the strategy we have adopted is the
following. It is generally accepted that the filaments occupy roughly 10\% of the observed
volume \citep{Forero-Romero:09,Jasche:09,Aragon-Calvo:10}. In the actual observed volume
$\nu(K)$ of the SDSS the observed filamentary network is made of (roughly
speaking, and with respect to the chosen unit measure), about $N_\mathrm{cyl}=3 \times 10^4$
cylinders. This gives a coarse cylinder density of about $N_\mathrm{cyl}/\nu(K)$ of
cylinders per unit of volume. Hence, a very intuitive way of fixing the range of
parameters is to equalise this density with the probability density given by the model of
having a cylinder in an observed location. This probability density is naturally
approximated by the conditional intensity of the model. Hence, we get
\begin{equation}
\frac{N_\mathrm{cyl}}{\nu(K)}=\lambda(\zeta ; \bmath{y}) = \frac{p(\bmath{y} \cap\{\zeta\})}{p(\bmath{y})}
\end{equation}
or
\begin{equation}
\log \left[\frac{N_\mathrm{cyl}}{\nu(K)}\right] = U(\bmath{y}) - U(\bmath{y} \cap\{\zeta\}),
\end{equation}
where $\zeta$ is the new cylinder to be added to the configuration $\bmath{y}$.

The definition of the data energy needs some predefined parameters. To test for the
``locally high density'', we fix $\rho_\mathrm{h}=1.0$: the assumed number density in a
filament should be at least three times larger than in its outer layers. For the ``locally uniform spread'' we set $\rho_\mathrm{u}=4$, this allows some lumpiness along the filament and at the same time penalises filaments that cross large clusters. The minimum
number of galaxies inside a cylinder is set to be $n_\mathrm{min}=3$. To assure balance
between the concentration and hypothesis testing terms, the constants in front of the hypothesis
testing term is chosen to be $c_\mathrm{hyp}\in[0.7,0.9]$. The value is chosen uniformly within the given interval. The distribution of the cylinder data potentials after the simulation
are shown in Fig.~\ref{fig_dataterm}, together with the distributions of the hypothesis
testing term and the cylinder concentration term: we see that for given data and chosen
parameters, these two terms are comparable and the overall data potential is reasonable.

For interaction energy, the potentials ($\log\gamma_k$, $\log\gamma_s$) are chosen from an
uniform prior density $p(\theta)$. We have opted for this choice since no information
concerning the relative strength is available \citep{Stoica:07a,Stoica:07,Stoica:10}.
Still, the general guidelines for fixing the prior parameters are that 2-connected
cylinders are generally encouraged, while 1-connected cylinders are slightly penalised and
0-connected cylinders are strongly penalised. This choice encourages the cylinders to
group in filaments in those regions where the data energy is good enough. Hence, the prior
domain was set to $\log\gamma_0\in[-2.0,-1.0]$, $\log\gamma_1\in[-1.0,0.0]$, and
$\log\gamma_2\in[0.8,1.8]$. The repulsion parameter $\gamma_k=0$, so configurations of
repulsing cylinders are forbidden. The prior domain for $\log\gamma_s$ was chosen based on
the distribution of data energies (see Fig.~\ref{fig_dataterm}): the chosen domain have to
be in balance (in the same range) as data potentials for cylinders.

In Table~\ref{tab:params} the parameters of the Metropolis-Hastings algorithm and the
simulated annealing algorithm are also given. For the change move, the maximum shift for
the cylinder centre is $\Delta k$ and the minimal cosine between the old and new direction
angles is $\Delta\omega \leq \lvert\omega_\mathrm{old}\cdot\omega_\mathrm{new}\rvert$. We
use a uniform prior for cylinder shift in a spherical volume with radius $\Delta k$:
orientations are taken uniformly on a unit sphere.

%=============================================================================
\subsection{Extracting the filamentary pattern spine}
\label{sect:spines}

\begin{figure}
	\includegraphics[width=84mm]{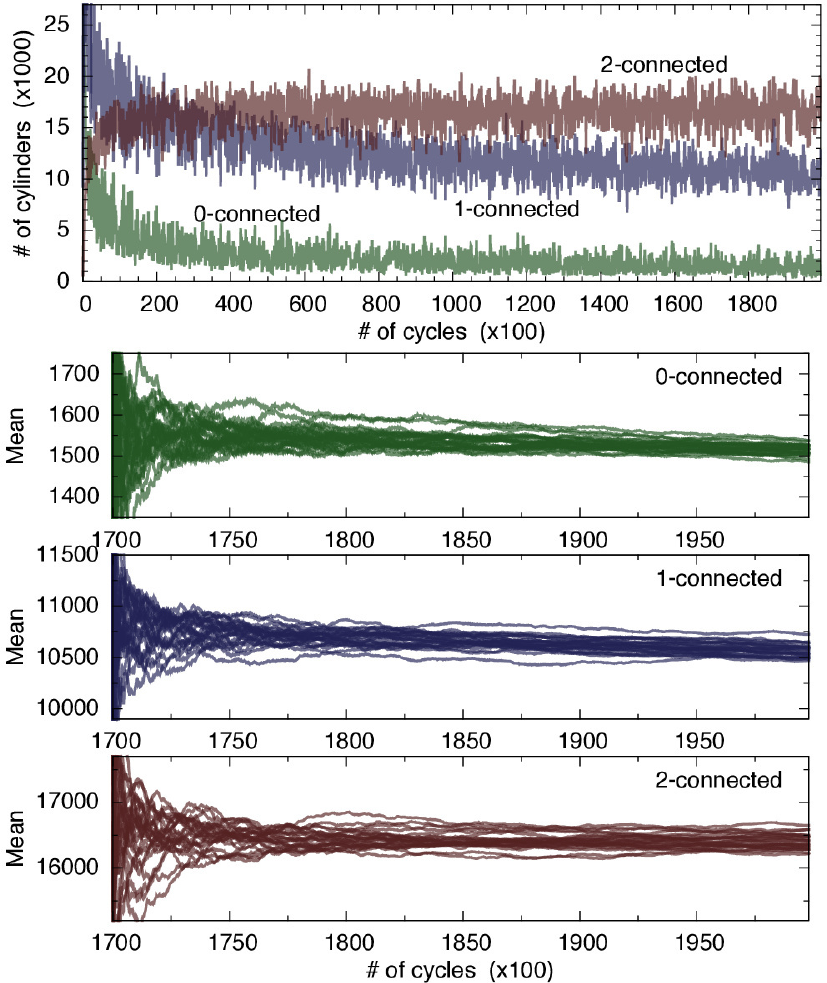}
    \caption{The time series of the numbers of cylinders (0,1,2-connected) in a configuration (\emph{upper 
	panel}). Cumulated means for 0,1,2-connected 
	cylinders (lower three 
	panels, respectively) computed for the final part of the simulation: superposition of 50 independent simulations.}
	\label{fig:nrcyl}
\end{figure}

\begin{figure}
	\includegraphics[width=84mm]{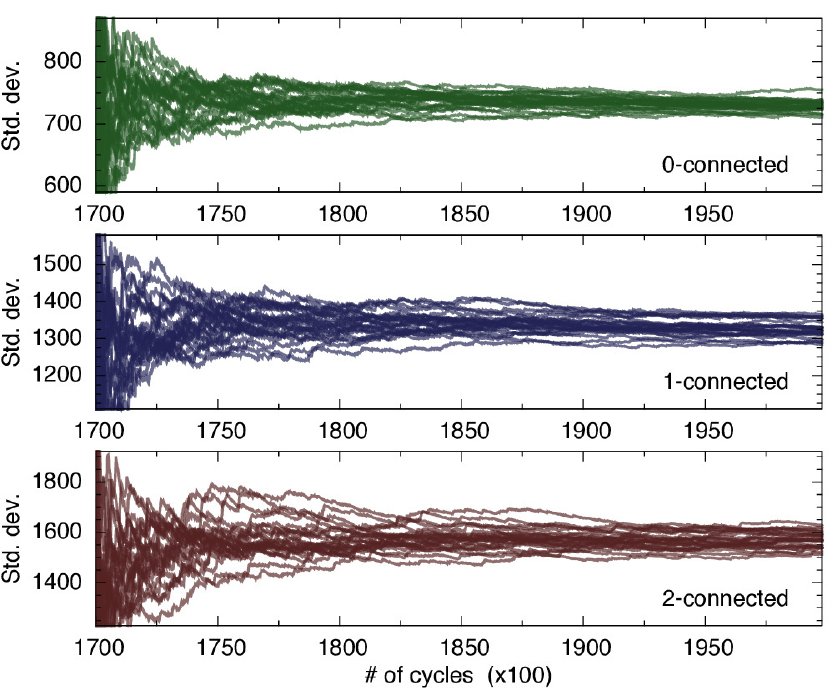}
    \caption{Cumulated standard deviations for 0,1,2-connected cylinders computed for the final part of the simulation: superposition of the 50 independent simulations.}
\label{fig:acu_stddev}
\end{figure}

\begin{figure}
\includegraphics[width=84mm]{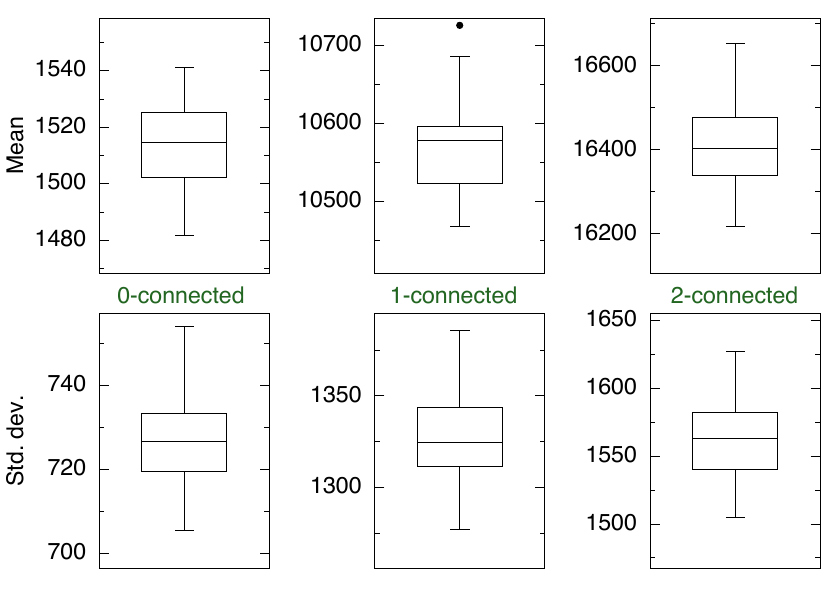}
\caption{Box-plots of the estimated mean (\emph{upper row}) and standard deviation (\emph{lower row}) distributions for 0,1,2-connected cylinders. The values are computed from the final configurations of 50 independent simulations.}
\label{fig:boxplot}
\end{figure}

The solution provided by our model is stochastic. Therefore some variation in the detected
patterns is expected for different runs of the method. In Fig.~\ref{fig:snapshot} (upper
left panel) a single MCMC simulation is showed, while indicating the different types of
cylinders: isolated (grey), single-connected (green), and double-connected (red). The
Fig.~\ref{fig:snapshot} (upper right panel) presents the superposition of $25$ independent
simulations. It can be seen, that the main features of the filamentary pattern are
detected by most of the simulations, while differences that appear are due to the random
effects of the method.

In order to have a more precise measure of these differences, a brief statistical
exploratory analysis was done. To do this the sufficient statistics of the model (the
number of 0,1,2-connected cylinders) were analysed. In one simulation, the simulated
annealing algorithm run during $20\times 10^9$ steps or moves. One step consists of one
iteration of the transition kernel of the MH dynamics, that is an accepted or rejected
birth or death or change proposal. The temperature was lowered every $10^5$ iterations (one cycle):
this number of moves was considered sufficiently high in order to obtain almost
un-correlated samples. Finally, we get $20\times 10^4$ samples per simulation. In all, we
have considered $50$ independent simulations.

The results of the exploratory analysis are shown in Figs.~\ref{fig:nrcyl}
and~\ref{fig:acu_stddev}. Figure~\ref{fig:nrcyl} (upper panel) shows the number of cylinders
in configuration as a function of cycles. We see that initially the number of
2-connected cylinders increases but it remains roughly constant after a certain time. The
number of 0- and 1-connected cylinders decreases: this decrease is expected since
simulated annealing penalises these cylinders more over time.
Eventually, these numbers also approach a constant value. The three lower panels in
Fig.~\ref{fig:nrcyl} show the cumulated means for the final part of the simulations:
superposition of 50 simulations are shown. From this figure we see that the number of
0,1,2-connected cylinders tend to have similar statistical values in every simulation. The cumulative standard deviations are shown in Fig.~\ref{fig:acu_stddev}. The box-plots of the mean and standard deviation distributions obtained from the final realisation of 50 independent simulations are shown in Fig.~\ref{fig:boxplot}. The standard deviation is much larger than the variation in mean numbers
of cylinders, showing that all the 50 simulations are statistically equivalent.

These numerical results are coherent with the detection obtained in Fig.~\ref{fig:snapshot}. The robust part of the network is given by the 2-connected cylinders. The part of the network made of 0- and 1-connected cylinders may be considered at a quick look as ``noisy''. Still, the question what part of this ``noisy'' part is relevant for the filamentary network, is of real interest. Our manner of answering it was to leave the model parameters rather free, since we do not know exactly how the objects we are looking for look like. Averaging several simulation results and spine detection should eliminate the ``noisy'' part while keeping the important short filaments.

There is another point to be outlined. Our plots show that simulated annealing does not reach convergence yet (in theory, it converges at infinity). This is due to the real computational time needed for getting the results: with modern computers, one simulation takes approximately 1000 CPU-hours in a single CPU.
There was a compromise to be done here: choosing an appropriate cooling schedule and stopping the algorithm  after a while, or choosing a fast cooling schedule and stating that the algorithm ``converged''. We have chosen the first approach. The convergence of the MCMC simulation methods is still an open research problem. For the reader
interested to get a deeper insight of this very interesting problem we recommend as a
starting point \citet{Robert:04}.

The main advantage of using such a stochastic approach consists in the ability of the
model to give a simultaneous morphological and statistical characterisation of the
pattern. This allows the comparison of several filamentary networks and this idea was used
in a previous study to compare mock catalogues and observed data \citep{Stoica:10}.
Nevertheless, it is legitimate to wish to have a smooth map of these filaments. A solution
to this problem is, under the hypothesis of the model, to estimate the probability that a
point belongs to the filamentary network and to look at those regions where these values
are higher than a given threshold \citep{Stoica:07a,Stoica:07,Stoica:10}. 
Several
realisations should be used to estimate this quantity and the filamentary network is
smoothed. In our previous work, these quantities are called visit maps, and this name is
kept in the following, due to its suggestivity. In mathematics, these quantities are
known as level sets and the convergence of these type of estimators was studied
in~\cite*{Stoica:12}. The visit map estimated from a number of $1000$ realisation is shown
in Fig.~\ref{fig:snapshot} (lower left panel).

Another very important aim of our work is to link the richness of our approach with the
very efficient existing deterministic methods for filaments finding. Therefore, in the following a method for filamentary
pattern spine detection is proposed. Figure~\ref{fig:snapshot} (lower right panel) shows
the result of the introduced method. The main difference with the existing methods is that
the spine detector we build uses the information provided by our stochastic approach. This
information consists of different quantities that can be estimated locally using our
model. These quantities estimate different probability and visit maps and also statistics
related to the orientation field induced by the filamentary network. 

The spine detection we propose is based on two main ideas. The first idea is that filament
spines are situated at the highest density regions outlined by the filament probability maps.
Next, in these regions of high probability for the filamentary network, the spines have an
orientation that is aligned with the direction given by the orientation field of the
filamentary network. On the contrary, in cross sections of filaments, the filament
detection probability can be high, but the orientation field is not clearly defined. The
filament spine detection method is described below in detail.

\begin{figure*}
	\includegraphics[width=88mm]{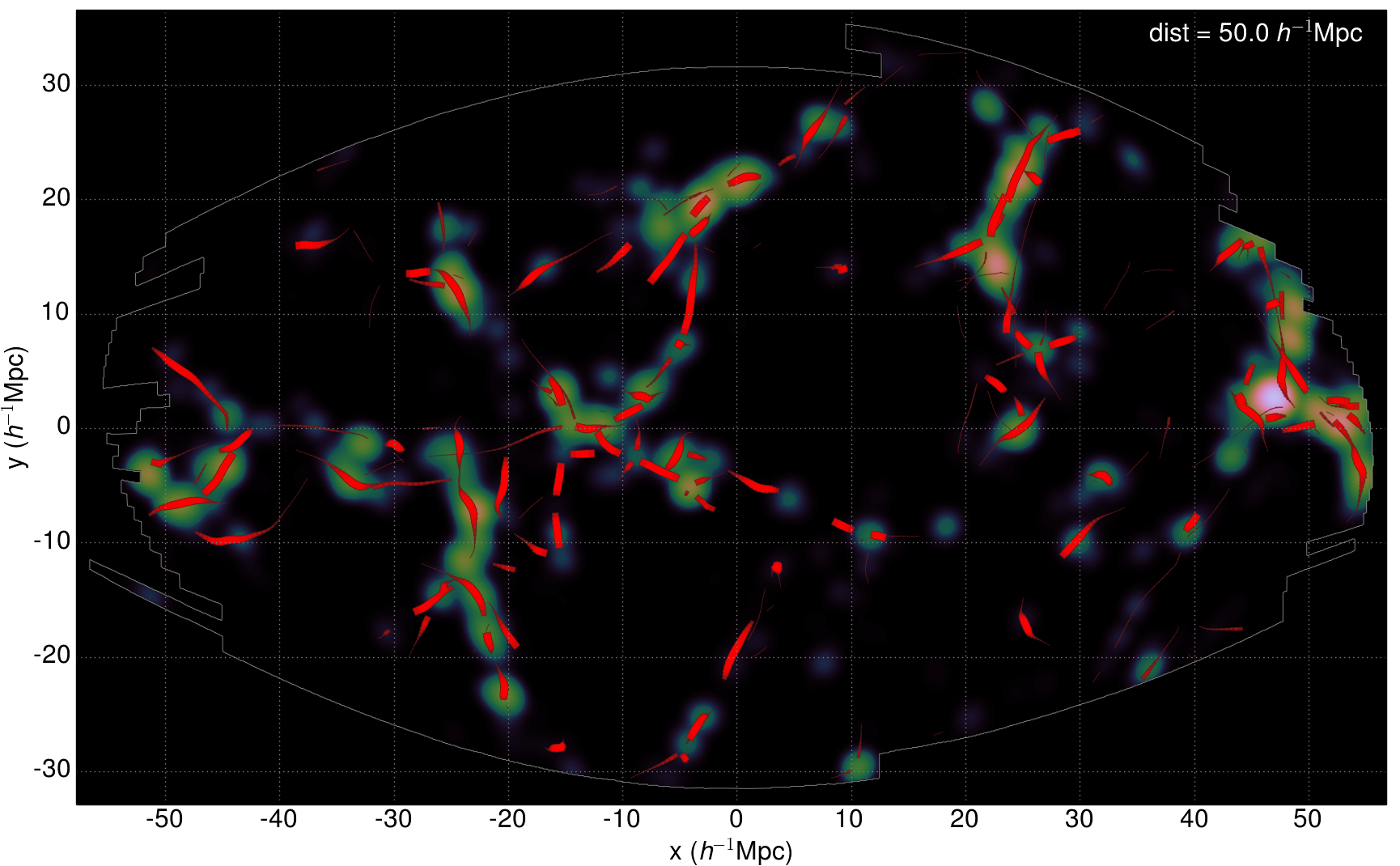}
	\includegraphics[width=88mm]{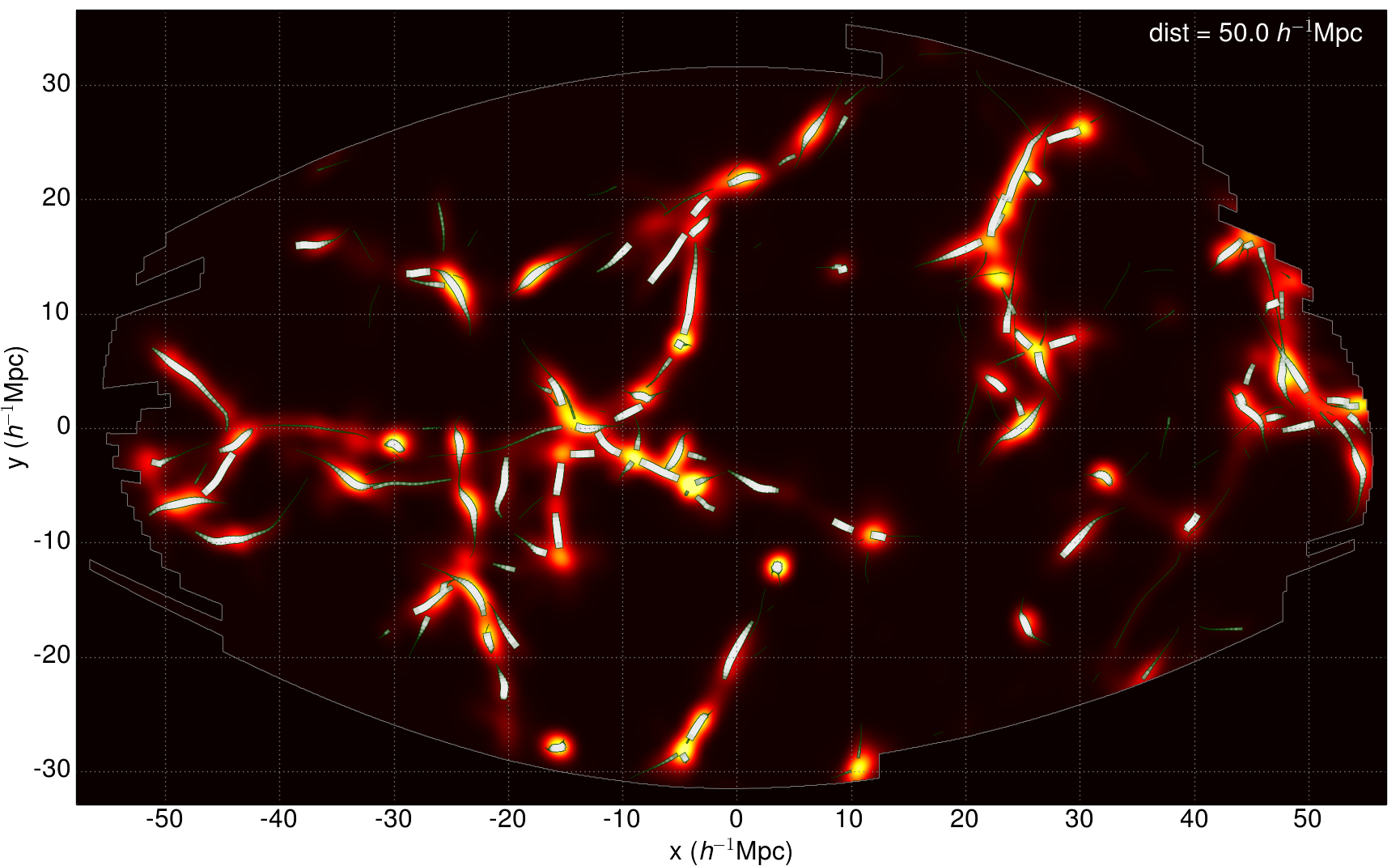}
	\includegraphics[width=88mm]{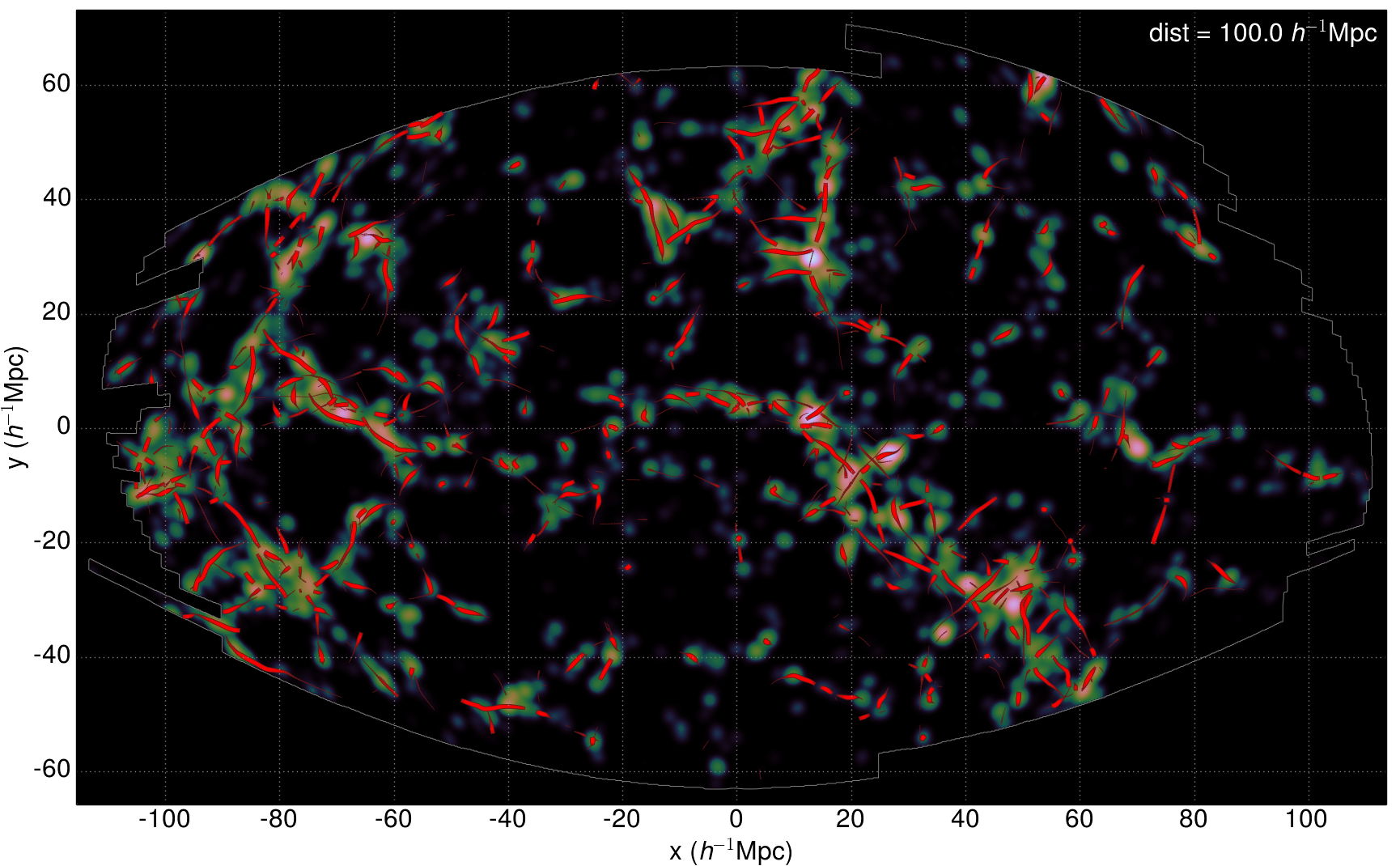}
	\includegraphics[width=88mm]{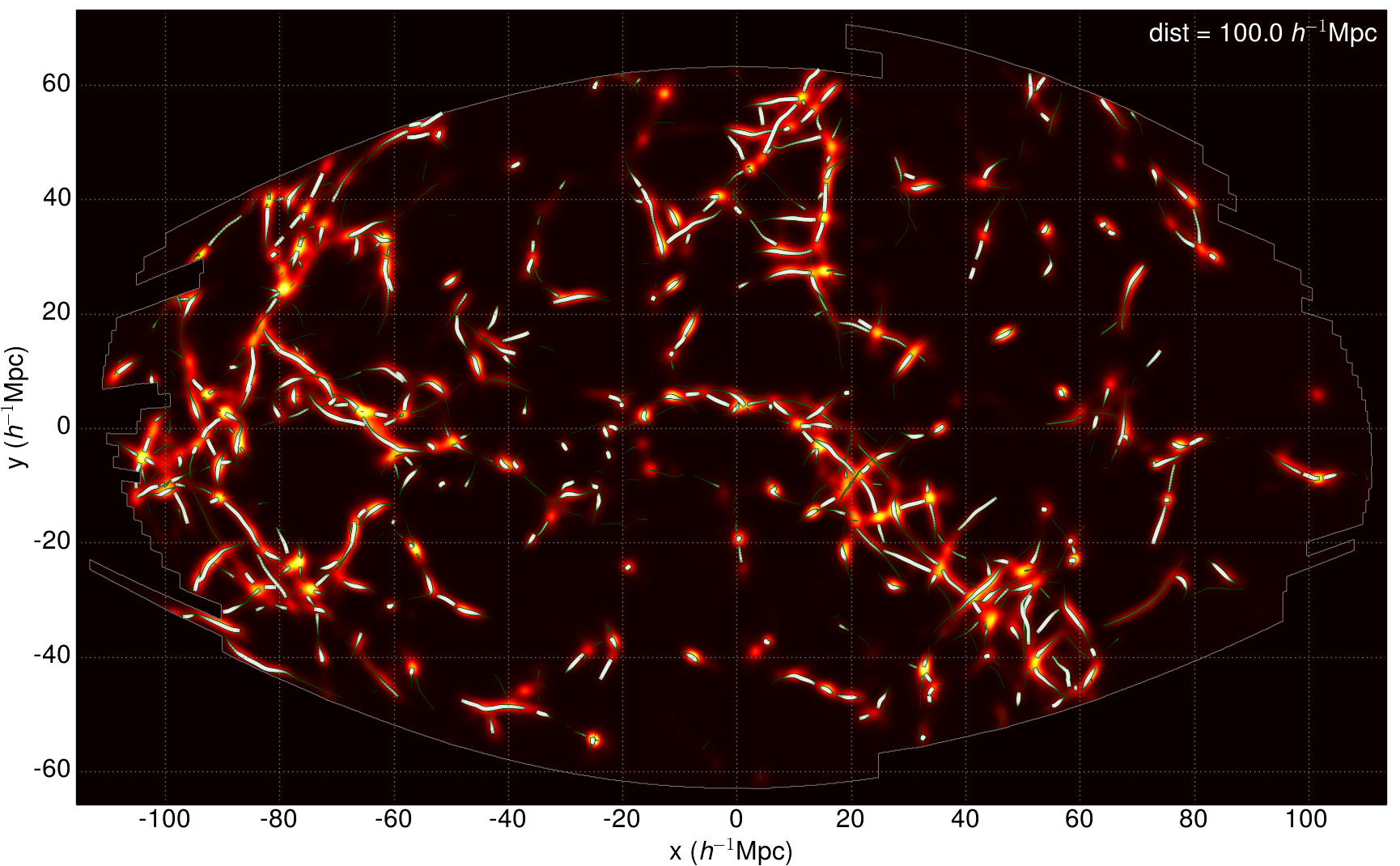}
    \caption{Sky projections of luminosity density field (left panels) and the visit map (right panels) at distances 50~$h^{-1}$Mpc (top row) and 100~$h^{-1}$Mpc
(bottom row). Luminosity density field is smoothed with 1~$h^{-1}$Mpc $B_3$ spline kernel. For better visualisation, both images are created by
summing up projected densities on several planes within range of $-4 \ldots +4~h^{-1}$Mpc from the indicated distance (using 1~$h^{-1}$Mpc step) and
presented in logarithmic scale. Extracted filaments in the same distance interval are drawn with red and white lines, the width of line
denotes the distance between filament and the plane of the image. There is good correspondence between the structures in the luminosity
density field and detected filaments. The fly-through movie, showing the full observed volume is available at
\url{http://www.aai.ee/~elmo/sdss-filaments/}. }
\label{fig:rmap}
\end{figure*}

First, we re-call the visit map estimator $\mathcal{L}(\bmath{k})$ for a given point
$\bmath{k}=(x,y,z)$
\begin{equation}
	\mathcal{L}(\bmath{k}) = \frac{1}{N}\sum\limits_{i=1}^N \mathbbm{1}\{\bmath{k} \in \bmath{Y}_i\},
\end{equation}
where $\bmath{Y}_1,\bmath{Y}_2,\ldots, \bmath{Y}_N$ are $N$ cylinder configurations and
$\mathbbm{1}\{\bmath{k}\in \bmath{Y}_i\}$ is the indicator function testing whether the point
$\bmath{k}$ belongs to any of the cylinders in the configuration $\bmath{Y}_i$. By this
definition, the visit map is defined to be in the range $\mathcal{L}(\bmath{k})\in [0,1]$.

The density map $\mathcal{D}(\bmath{k})$ of filaments is defined as a weighted visit map
(level set). For a given point $\bmath{k}=(x,y,z)$ it is defined as
\begin{equation}
	\mathcal{D}(\bmath{k}) = \frac{1}{N}\sum\limits_{i=1}^N \frac{ \sum\limits_{y \in\bmath{Y}_i} \exp\left[v(y)\right] \mathbbm{1}\{\bmath{k}\in y\} }
	{\sum\limits_{y\in\bmath{Y}_i} \mathbbm{1}\{\bmath{k}\in y\}},
\end{equation}
where the first summation is over realisations and the second summation is over cylinders
in that given configuration $\bmath{Y}_i$. The potential function $v(y)$ for a cylinder $y$ is
defined by Eq.~(\ref{eq:datapotential}). The indicator function $\mathbbm{1}\{\bmath{k}\in
y\}$ acts as a cylindrical kernel and selects the points that the cylinder $y$ covers.
Note that a point $\bmath{k}$ can be covered by a several cylinders in one configuration,
but effectively all configuration have equal weights. We weight the visit map to suppress
weak intersecting filaments (to reduce the noise) and to encourage stronger filaments.

The orientation field $\mathcal{G}(\bmath{k},\bmath{\omega})$ for a point $\bmath{k}$ and
for a orientation $\bmath{\omega}=\phi(\eta,\tau)$ is defined as
\begin{equation}
	\mathcal{G}(\bmath{k},\bmath{\omega}) =
	\frac{ \sum\limits_{i=1}^N  \sum\limits_{y\in\bmath{Y}_i}\exp\left[v(y)\right]
	 \mathbbm{1}\{\bmath{k}\in y\} \lvert \bmath{\omega}\cdot\omega_y \rvert}
	{\sum\limits_{i=1}^N 
	\sum\limits_{y\in\bmath{Y}_i}\exp\left[v(y)\right]\mathbbm{1}\{\bmath{k}\in y\}},
\end{equation}
where $\bmath{\omega}\cdot\omega_y$ denotes the scalar product between the orientation
vector $\bmath{\omega}$ and the cylinder orientation $\omega_y$. Using this definition,
$\mathcal{G(\bmath{k},\bmath{\omega})}\in [0,1]$. Using the orientation field, we also
define density field for orientation strengths: the maximum of the orientation field
depending on $\bmath{\omega}$ at a given location $\bmath{k}$ is
$\mathcal{D_G}(\bmath{k})$,
\begin{equation}
	\mathcal{D_G}(\bmath{k}) = \max\left\{ \mathcal{G}(\bmath{k},\bmath{\omega}) \right\}.
\end{equation}
This quantity is a weighted estimator of the expectation of the scalar product between the
orientation $\bmath{\omega}$ at the location $\bmath{k}$. If the cylinder orientation $\omega_y$ is uniform
on the unit sphere, then the absolute values of the scalar product is a uniform random
variable between $0$ and $1$. Hence the value of $\mathcal{D_G}$ under the uniform
assumption should be close to $0.5$. If all the cylinders are aligned with respect to
$\bmath{\omega}$ then the value of $\mathcal{D_G}$ should be close to $1$. If we are interested in
the situation that the majority of cylinders are aligned to $\bmath{\omega}$, then we may test
$\mathcal{D_G} > \mathcal{D_G}_\mathrm{lim}$ with $\mathcal{D_G}_\mathrm{lim}$ a pre-defined threshold value rather close
to $1$. For our purposes we set $\mathcal{D_G}_\mathrm{lim}=0.75$, since this value may suggest a half way
distance between the uniform and the completely aligned case.

The corresponding orientation of the maximum value $\mathcal{D_G}(\bmath{k})$ at location $\bmath{k}$ is $\bmath{\omega}_{\mathcal{G}}(\bmath{k})$ and it is defined as
\begin{equation}
	\bmath{\omega}_{\mathcal{G}}(\bmath{k}) = \arg \max\limits_{\bmath{\omega}}\left\{ \mathcal{G}(\bmath{k},\bmath{\omega}) \right\}.
\end{equation}

For computing the previous estimators the last $32$ extracted realisation from a single run of the
algorithm were kept: we extracted the realisations after 1000 cycles. In total we have used $50$ independent runs of the algorithm. This
gives in all $1600$ cylinder configurations to be used for computing the previous defined
quantities.

At a first look, the previous quantities can be computed locally, hence there is no need for
keeping track of the cylinder configurations. In our case we need
to calculate the visit map (and orientation map with orientations) in a grid with
grid-step smaller than 0.1~$h^{-1}$Mpc to accurately determine the spine of the filaments.
Due to the limitations of the computer memory, it cannot be computed globally for the
entire simulation box. In order to calculate the visit map and orientation map with
sufficiently fine grid, they have to be computed locally. For that purposes, we store the
cylinder configurations from every simulations and compute the visit maps and orientation
maps locally as defined above. The advantage of this approach is that all required
quantities can be computed for every space and orientation in the sample volume and we are
free of gridding. The chosen 1600 cylinder configuration is large enough for visit map and
orientation map estimation and at the same time it requires reasonable amount of
computational resources.

Using previous definitions, for every point $\bmath{k}$ we have three values: the filament
density $\mathcal{D}(\bmath{k})$, the orientation strength $\mathcal{D_G}(\bmath{k})$, and
the filament orientation in that location $\bmath{\omega}_{\mathcal{G}}(\bmath{k})$. To
extract a single filament using these three quantities, we do the following.
\begin{enumerate}
	\item We start at a  point of the highest density $\mathcal{D}(\bmath{k})$ that is not 
	yet masked out
	(we will discuss masking later). We designate this point as $\bmath{k}_0$. The initial 
	density map is calculated on a 0.5~$h^{-1}$Mpc grid, which is sufficiently fine 
	(compared with cylinder size) for global maxima. After maximum is found, the density 
	map is calculated locally on a 0.01~$h^{-1}$Mpc grid. Initially, all the regions where 
	$\mathcal{L}(\bmath{k})<\mathcal{L}_\mathrm{lim}=0.05$ are masked out: e.g. we are 
	searching for the filaments in the regions that have been covered at least in 5\% of 
	the realisations. We remind that all the detected structures are filaments by 
	definition, and this is only the detection probability that depends on the model 
	parameters.
	\item If the orientation at that point is defined, we start extracting a filament. We 
	say that the orientation is defined if 
	$\mathcal{D_G}(\bmath{k}_0)>\mathcal{D_G}_\mathrm{lim}=0.75$. Otherwise, we mask out the region 
	around this point and continue with the step (i). The size of the masked region is 
	taken as $1.0\,h^{-1}\mathrm{Mpc}$.
	\item We look from the point $\bmath{k}_0$ to both sides along 
	$\bmath{\omega}_{\mathcal{G}}(\bmath{k}_0)$.
	\item To extract the filament, we move from the point $\bmath{k}_0$ in the direction
	 $\bmath{\omega}_{\mathcal{G}}(\bmath{k}_0)$ 
	 by $\delta x=0.5\,h^{-1}\mathrm{Mpc}$. The step size is arbitrary, but a smaller step 
	 size gives smoother filaments. The step size $\delta x = r$ is good enough ($r$ is 
	 the cylinder radius). We designate the new point as $\bmath{k}_i$.
	\item We calculate the density map 
	$\mathcal{D}_{\bmath{\omega}_{\mathcal{G}}}(\bmath{k}_i)$ that is perpendicular to the 
	direction $\bmath{\omega}_{\mathcal{G}}(\bmath{k}_i)$. The density map 
	$\mathcal{D}_{\bmath{\omega}_{\mathcal{G}}}(\bmath{k}_i)$ is two-dimensional and from 
	that map we find the maximum that is closest to the point $\bmath{k}_i$: the location 
	of this maximum is marked as $\bmath{k}_{i\prime}$.
	\item If the orientation is not defined at $\bmath{k}_{i\prime}$, we stop the filament 
	extracting algorithm and continue with the step~(x).
	\item If the orientation is defined 
	($\mathcal{D_G}(\bmath{k}_{i\prime})>\mathcal{D_G}_\mathrm{lim}$) we go forward by 
	$\delta x$ and find a new point as previously explained. This point is needed to 
	perform two additional checks.
	\item Firstly, to avoid breaks in the filament, we calculate the curvature of the 
	filament at the point $\bmath{k}_{i\prime}$ using this point and its neighbours. The 
	curvature $\kappa=1/R$, where $R$ is the radius of the sphere that these three points 
	touch. We use the limiting value $\kappa>\kappa_\mathrm{lim}=1.0$ to stop the filament 
	finding algorithm.
	\item Secondly, we require that the orientation at the point $\bmath{k}_{i\prime}$ and 
	at the neighbouring points is roughly the same:
	\begin{equation}
		\max\lvert  \bmath{\omega}_{\mathcal{G}}(\bmath{k}_{i\prime})\cdot 
		\bmath{\omega}_{\mathcal{G}}(\bmath{k}_{i\pm 1}) \rvert
		>\tau_\mathrm{lim}=0.95.
	\end{equation}
	If the tests are not satisfied, we stop the filament finding algorithm. Otherwise, we 
	move in the direction $\bmath{\omega}_{\mathcal{G}}(\bmath{k}_{i\prime})$ by 
	$\delta x$ and continue with the step (v).
	\item If all the filament points from both sides of $\bmath{k}_0$ have been found, we 
	mask out the region that this filament covers. The radius of the masked out region is 
	taken $1.0\,h^{-1}\mathrm{Mpc}$ (twice the filament radius). We save the extracted 
	points as a single filament.
	\item We return to the point (i) until all the volume is masked out.
\end{enumerate}

Basically, this algorithm walks along the mountain chain in the filament density map and
tests if the orientation is defined and the orientation is the same as the walking
direction. There are only four parameters that define the filaments:
$\mathcal{L}_\mathrm{lim}=0.05$ defines the limiting visit map density and the strength of
a filament, $\mathcal{D_G}_\mathrm{lim}=0.75$ defines the orientation and estimates the
strength of orientation for a filament, $\kappa_\mathrm{lim}=1.0$ defines the limiting
curvature, and $\tau_\mathrm{lim}=0.95$ defines the angle between the filament and the
orientation field. All these criteria are unimportant for strong filaments, but they
influence the regions where filaments intersect or the regions where filaments are poorly
defined.

%=============================================================================
\section{Results and discussion}
\label{sect:results}

Figure~\ref{fig:rmap} illustrates the detected filaments and their axes. In left panel are
shown the luminosity density field smoothed with 1~$h^{-1}$Mpc $B_3$-spline kernel. Right
panel shows the corresponding visit map ($\mathcal{L}$) for detected filaments. Extracted filament spines are shown with lines.
Qualitatively, the filament axes plotted in these figures appear to closely trace the
underlying large-scale filaments. This is not surprising because, by definition, our
filament finder is based on cylindrical shapes and the filament detection probability
field should trace the filamentary structures. \citet{Tempel:13} used the Bisous process
to detect the filaments from dark matter simulations and showed that the detected
filaments are very well aligned with the underlying velocity field. This shows that the detected filaments are also dynamical structures.

\begin{figure}
	\includegraphics[width=84mm]{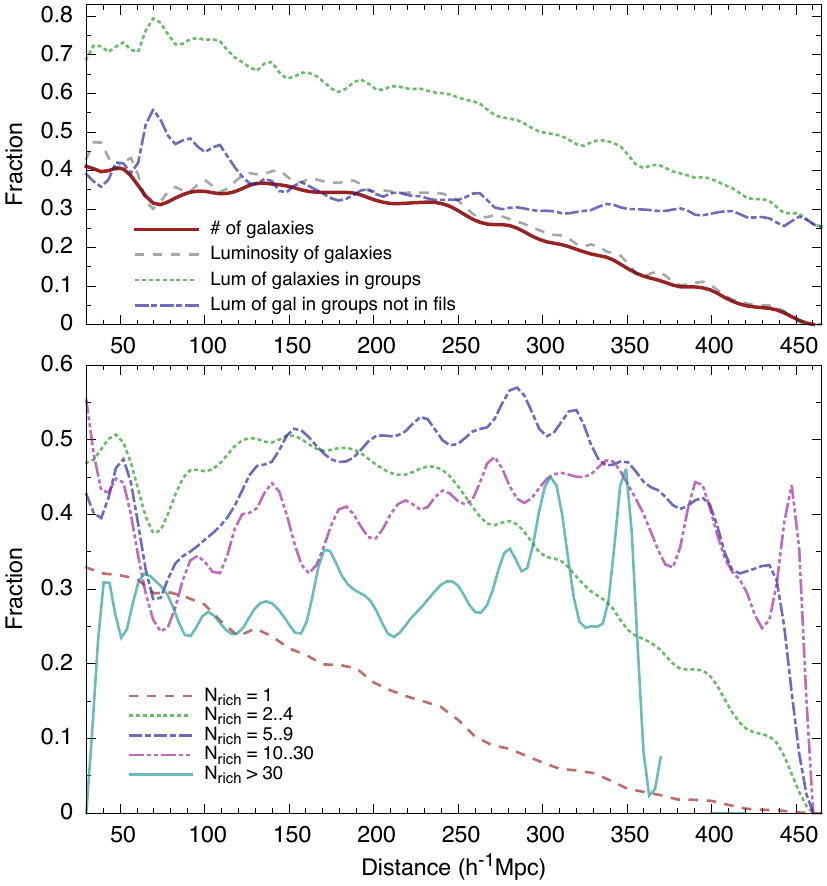}
    \caption{The upper panel shows the fraction of galaxies in filaments (red solid line) 
	and the fraction of observed luminosity in filaments (grey dashed line) as a function 
	of distance. The green dotted line shows the fraction of luminosity in groups and the 
	blue dot-dashed line shows the fraction of luminosity in groups that are not in 
	filaments. The filament radius is taken to be $0.5~h^{-1}\mathrm{Mpc}$. The lower 
	panel shows the fraction of galaxies in filaments for different group richness bins.}
	\label{fig:fraction}
\end{figure}

The upper panel in Fig.~\ref{fig:fraction} shows the fraction of galaxies in filaments as
a function of distance: the red solid line shows the number of galaxies and the grey
dashed line shows the fraction of luminosity in filaments. We note that there is slightly
more luminosity in filaments than the number fraction predicts -- meaning that in
filaments the luminosity density of galaxies is higher than in average. The green dotted
line in the upper panel of Fig.~\ref{fig:fraction} shows the fraction of luminosity that
is in groups \citep[we use the groups as defined in][]{Tempel:12}. Since we use a
flux-limited sample, the number density of groups (and their luminosity) is higher for
nearby regions. The blue dot-dashed line shows the fraction of luminosity in groups that
are not in filaments. We see that this fraction is almost constant. The lower panel
in Fig.~\ref{fig:fraction} gives an explanation for that. Most of the galaxies that are in
filaments are also in small groups and since filaments have a chain-like inner structure,
the nearby filaments are made of smaller groups that are aligned. Further away, the number
density of smaller groups is lower and the filament detection probability is also lower
(the red solid line in the upper panel). The lower panel in Fig.~\ref{fig:fraction} also
shows that isolated galaxies are preferentially not located in filaments and also the
galaxies in large clusters are mostly not in filaments: only galaxies in the outskirts of
large clusters are in filaments.

\begin{figure}
	\includegraphics[width=84mm]{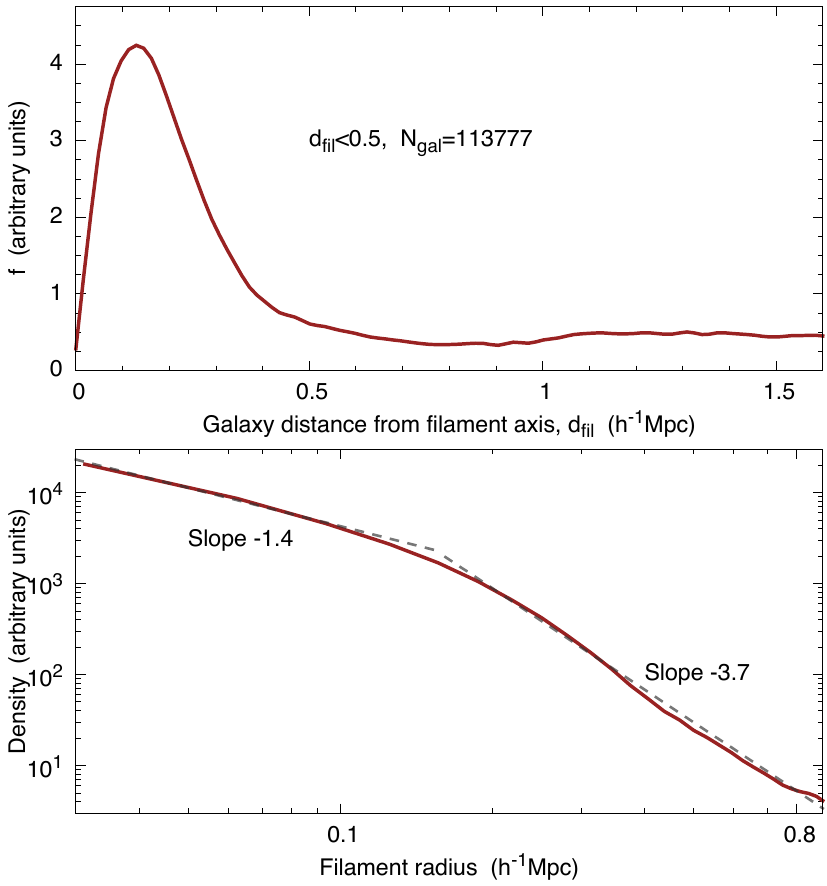}
    \caption{Upper panel: the distribution of galaxy distances from the nearest filament 
	axis. After $0.5~h^{-1}\mathrm{Mpc}$, the distribution is dropping since our defined 
	filament radius is $0.5~h^{-1}\mathrm{Mpc}$. The number of galaxies closer than 
	$0.5~h^{-1}\mathrm{Mpc}$ to the filament axis is shown in the figure. Lower panel: the 
	number density of galaxies as a function of the distance from the axis.}
	\label{fig:rad_distr}
\end{figure}

All this implies that filaments are far from being smooth uniform structures. Visual
inspection of the density field and the spines of filaments (Fig.~\ref{fig:rmap}) shows
that filaments are populated by small galaxy groups and large clusters are in intersection
of those filaments, as already shown in \citet{Bond:96}. The same impression is
quantitatively confirmed in Fig.~\ref{fig:fraction}.

The fraction of galaxies (or luminosity) in filaments gives us roughly the mass filling
fraction of filaments. Up to distance $250~h^{-1}\mathrm{Mpc}$, the fraction is 35--40\%,
which is in very good agreement with $N$-body simulations
\citep{Forero-Romero:09,Aragon-Calvo:10,Hoffman:12}. Using the SDSS data, the same
filament mass filling fraction has been measured by \citet{Jasche:09}. We note that after
$250~h^{-1}\mathrm{Mpc}$, the number of detected structures decreases. \citet{Smith:12}
search structures in the SDSS in the same scale as we, and their number of detected structures
follows the same behaviour with distance. This is logically expected, since the number
density of objects (that decreases with distance in flux-limited survey) is strictly
related to the number of detected structures.

Although filaments have been studied extensively in general, there are only few
studies addressing their radial density profile
\citep{Colberg:05,Dolag:06,Aragon-Calvo:10}. However, in these papers, larger filaments
are considered, so a strict comparison is not possible. Figure~\ref{fig:rad_distr} shows
the filament radial profile in the current study: the upper panel shows the
distribution of the number of galaxies per radius, the lower panel shows the number
density profile for filaments. We see that most of the galaxies in filaments are closer
than $0.5~h^{-1}$Mpc to the filament axis. This is because our defined filament radius is
$0.5~h^{-1}$Mpc. We also see that there is a break in the density profile around
$r=0.2~h^{-1}$Mpc. Using weak lensing, \citet{Jauzac:12} studied a single filament that
connects two clusters and they also see the break in the density profile roughly at the
same distance. Direct comparison with other studies is possible, once we extract thicker
filaments: this is planned for future work.

\begin{figure}
	\includegraphics[width=84mm]{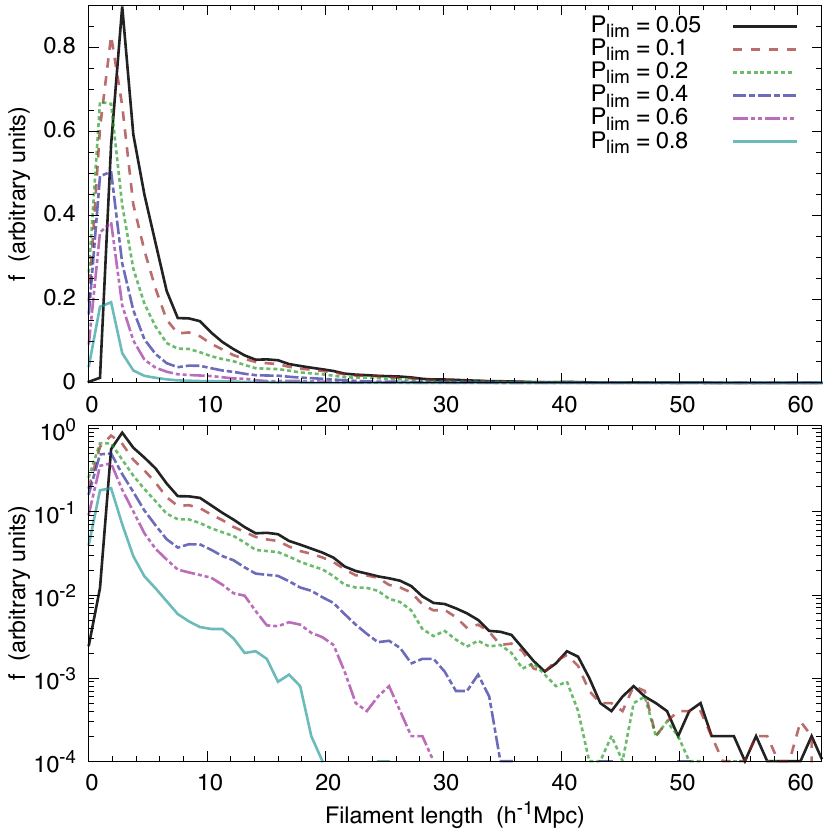}
    \caption{The filament volume distribution as a function of the filament length. The 
	distribution is shown for various limiting covering probability values ($\mathcal{L}_\mathrm{lim}$). The lower 
	panel emphasises the differences for long filaments.}
	\label{fig:length_distr}
\end{figure}

Another interesting quantity is the filament length distribution.
Figure~\ref{fig:length_distr} shows the filament length distribution in the linear (upper
panel) and logarithmic (lower panel) scale. The distribution is shown for different
minimum detection probability values ($\mathcal{L}_\mathrm{lim}$). The black solid line shows the filaments as given in
our catalogue. Increasing the limiting detection probability, the filaments start to
fragment and small (weak) filaments disappear. We see that increasing the detection
probability up to 0.2 practically does not change the long-end of the distribution. The
longest filaments are strong and dominant filaments. We emphasise that the value of the
detection probability is model dependent and it does not reflect the probability of the
filamentary structure (all detected structures in our model are filaments based on the
definition).

The filament length distribution has been also studied in \citet{Bond:10} using $N$-body
simulations. Compared with our length distribution, the longest filaments are roughly the
same, but we have more short filaments. This is probably because the spatial distribution
of observed galaxies is sparse and in many cases we only found a piece of a filament.

The maximum length of our filaments is $\sim 60~h^{-1}\mathrm{Mpc}$. This is in very good
agreement with other measured values \citep*{Bharadwaj:04,Bond:10,Pandey:11}.

\begin{figure}
	\includegraphics[width=84mm]{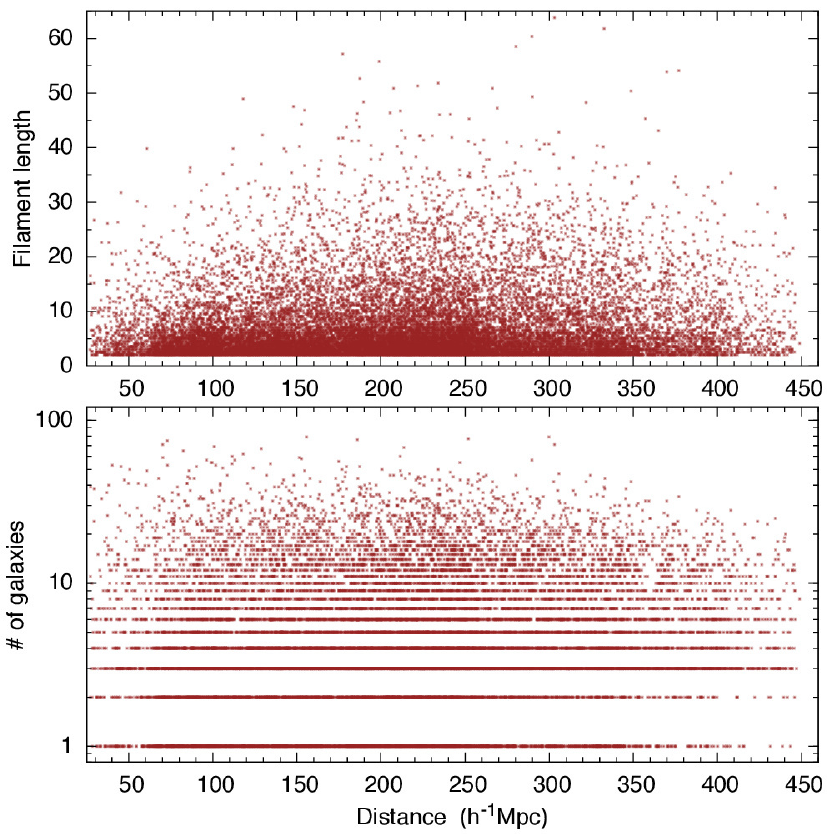}
    \caption{\emph{The upper panel} shows the filament length as a function of distance from the observer. 
	\emph{The lower panel} shows the number of galaxies in a filament as a function of 
	distance.}
	\label{fig:dist_par}
\end{figure}

Figure~\ref{fig:dist_par} shows the filament lengths and the number of galaxies in
filaments as a function of distance. We note that both distributions are quite uniform.
There is lack of long filaments in the nearby region because its volume is small. Further
away, the longest filaments are missing because the number density of galaxies is too low.
However, there exist filaments with lengths up to $30~h^{-1}$Mpc further than
$400~h^{-1}$Mpc. The number of galaxies in filaments (lower panel) decreases with distance
because we used a flux-limited sample: the faintest galaxies are missing further away.

\begin{figure}
	\includegraphics[width=84mm]{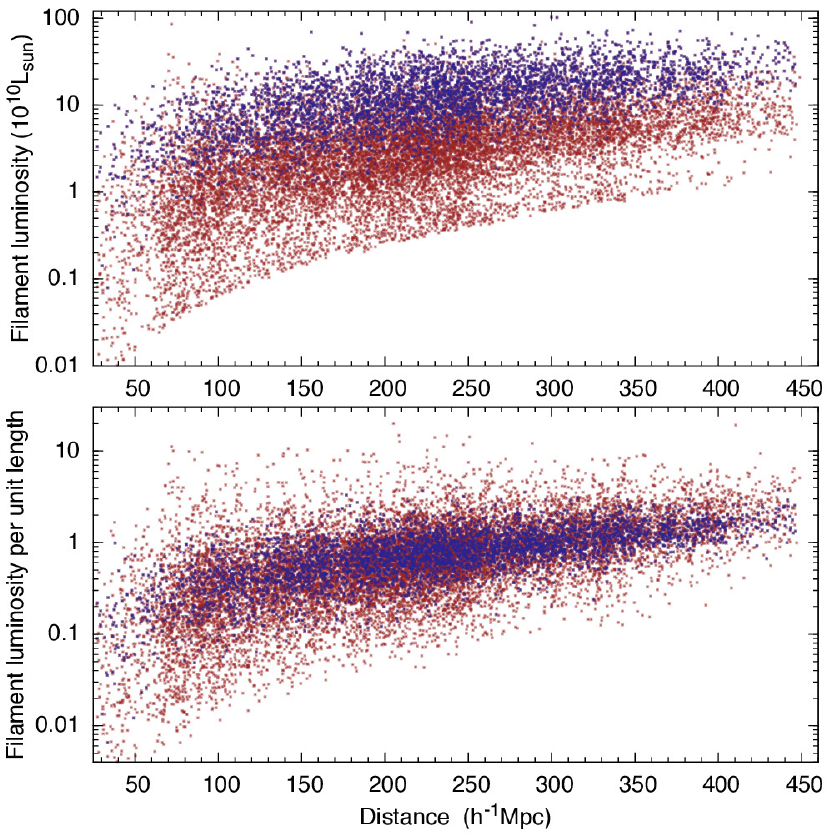}
    \caption{\emph{The upper panel} shows the filament luminosity as a function of 
	distance. \emph{The lower panel} shows the change of filament luminosity per unit 
	length with distance. The distribution for the luminosity per unit length is much more 
	tight. Red dots are for all filaments, blue dots are for long filaments (at least 
	$10~h^{-1}$Mpc long).}
	\label{fig:dist_lum}
\end{figure}

Figure~\ref{fig:dist_lum} shows the luminosity of a filament (upper panel) and the luminosity
per unit length (lower panel) as a function of distance. We notice that the faintest
filaments are missing further away, because of the flux-limited survey. However, the upper
limit is distance independent and it shows that the brightest filaments nearby and further
away are practically the same. We also note that the scatter in the lower panel is quite
small, indicating that luminosity per unit length in filaments does not vary much. Blue
points in the figure show the longest filaments (at least $10~h^{-1}$Mpc long). We see that
the longest filaments are also the most luminous filaments, however, their luminosity per
unit length lies within the average. This indicates that short and long filaments have on
average the same luminosity density.

\begin{figure}
	\includegraphics[width=84mm]{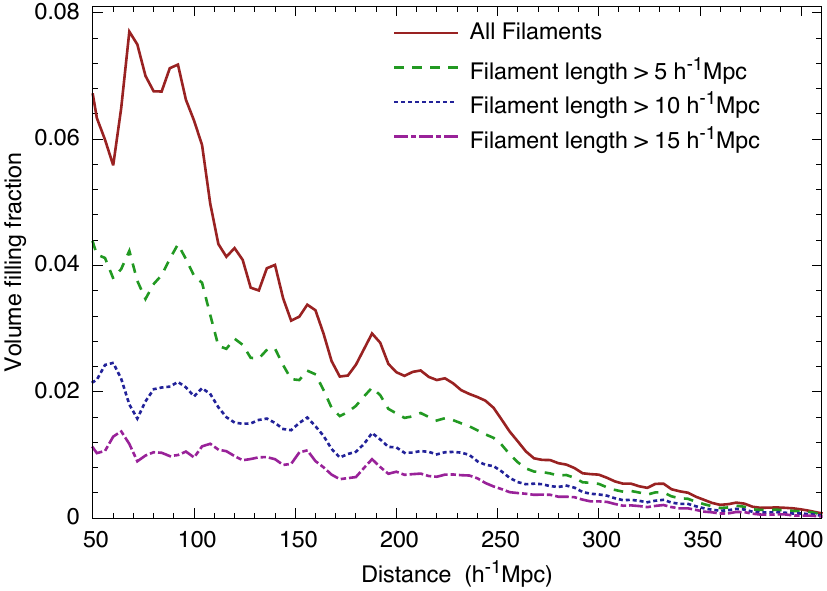}
    \caption{The volume filling fraction of filaments as a function of distance. Red solid line shows the volume filling fraction for all detected filaments. Other lines show the volume filling fraction, when only longer filaments are considered. The filament volume is calculated using the spines of the filaments and the filament radius is taken to be $r=1\,h^{-1}\mathrm{Mpc}$.}
	\label{fig:vff_dens}
\end{figure}

One important quantity that describes filaments is their volume filling factor.
Figure~\ref{fig:vff_dens} shows the volume filling factor as a function of distance. Since
the number density of galaxies decreases with distance, the volume filling factor also
decreases. The filament volume is calculated around the detected spines, using the filament radius $r=1.0~h^{-1}\mathrm{Mpc}$; in the
nearby region, the filling factor is $\sim 7\%$ and it decreases with distance (due to the
flux-limited survey). From $N$-body simulations, the volume filling factor has been
measured by \citet{Forero-Romero:09} and \citet{Aragon-Calvo:10}. In these papers, the
volume filling factor depends on the used threshold, but it is in order of 10\%, which is
in good agreement with our results. Based on the SDSS data, the filaments have been
extracted by \citet{Jasche:09}. They use a novel Bayesian sampling algorithm, which
permits precise recovery of poorly sampled objects in a non-linear density field. Based on
their analysis, the filament volume filling factor is 10--20\%, which is slightly larger
than ours. However, since they use a $3~h^{-1}\mathrm{Mpc}$ grid their filaments are
thicker than ours and these results cannot be directly compared. On the contrary,
\citet{Hoffman:12} showed that the filament volume filling fraction is 4--5\% using the
velocity shear tensor. This is slightly lower than we found in the current study
(considering the fact that we do not detect all structures). Since the mass filling
fractions in both studies are comparable, the filaments found in the velocity field are
probably located mostly in higher density environments and are somewhat thicker
\citep{Tempel:13}.

\subsection{Robustness of the filamentary pattern detection}

Our method is sensitive to the galaxy density. If a data set contains too few galaxies, no filaments will be detected since the filamentary pattern is not observable. If the data set contains many galaxies, the filamentary network is already defined, hence our method will always work. Indeed, intuitively, there is an optimal range for the galaxy number density so that our model delineates correctly the filaments.

In general, the model parameters depend on the minimum number density.
The model parameters of our ``machinery'' were designed studying the SDSS data set. After several trials and errors, we found the parameter values that give the best results (see Table~\ref{tab:params}). To reliably determine filaments, two (preferentially three or more) cylinders have to be aligned and connected. With the present model parameters, this leads to the minimum number density (inside a filament) of 6 galaxies within a cylindrical volume of the radius 0.5 and the length 6--10~$h^{-1}$Mpc. 

As the data energy in our model is determined by the ratio of densities in the
cylinder and its shadow, it does not depend on the local number density of 
galaxies (for fixed model parameters, and if the minimum number density condition is satisfied). This allows us to detect physically similar filaments regardless of the environmental density, and our method can recover structures of relatively sparsely sampled objects (filaments in lower density environments).

In a flux-limited survey, the number density of galaxies decreases with distance. For our model, it means that the filament detection probability decreases (we do not detect all filaments further away and/or we only detect parts of the filamentary network), but the reliability of the detected filaments,  determined by the visit map value, is largely unaffected, due to the robustness of the model.

As an example, Figure~\ref{fig:vff_dens} shows the volume filling fraction of filaments as a function of distance. Including  all filaments (red line), the sample is not homogeneous. However, it is possible to construct a statistically homogeneous sub-sample of filaments, when the sample is limited by distance and filament length. Figure~\ref{fig:vff_dens} shows that when using only longer filaments, the volume filling fraction is roughly constant with distance up to 240~$h^{-1}$Mpc. Further away, the galaxy number density decreases rapidly, and the full filamentary network for the scale used here ($r=0.5\,h^{-1}\mathrm{Mpc}$) is not clearly outlined.

Another important point to be raised is the question whether the optimal choice of the model parameters should depend on the galaxy number density. \citet{Stoica:10} addressed this issue and concluded that simple choices (e.g. increasing the cylinder size) do not produce good results: different cylinder sizes detect different structures. To reduce the incompleteness in filament detection, larger cylinders should be used everywhere to detect the same structures.

How can shot noise affect the filament detection in low number density regions? Like for any other methods, shot noise affects the results. The main advantage of our method is in its probabilistic nature. Individual realisations of the solution may be sensitive to noise. Still, averaging these realisations reduces the noise influence, and allows computation of robust statistical quantities \citep{Stoica:10}. These quantities are the estimates of the sufficient statistics, the level sets (visit maps) and the local detection probability.

It is also possible to think about the reverse formulation of the detection problem. That is, knowing the topological structure of the filamentary pattern, we may wonder what is the density range within the observed volume, that still outlines the given filamentary network. However, the optimal model choice for filament detection, is at the moment an open mathematical and data analysis problem.

%=============================================================================
\section{Conclusions and future work}
\label{sect:discussion}

This paper uses and develops an object point process with interactions (the Bisous
process) to trace the filamentary network in the flux-limited SDSS data. This method works
directly on the galaxy distribution and does not require any additional smoothing, it only
requires fixing the scale of structures. For the current work, we fixed the radius of a
filament as $r=0.5~h^{-1}\mathrm{Mpc}$, which is close to the scale of galaxy
groups/clusters; such filaments should have the largest impact for galaxy formation and
evolution.

Our filament finder is probabilistic in the sense that it gives us the filament detection
probability field together with the filament orientation field. Using these two fields, we
define the spines of the filaments and extract single filaments from the data. We showed
that the detected filaments fit well with the visible large-scale structure. The composed
catalogue of filaments for the SDSS is made publicly available (see Appendix~\ref{app:1}).

We showed that the filament mass and volume filling factors are in good agreement with
structures found in $N$-body simulations and in previous observational studies. The mass
filling fraction of our filaments is 35--40\%, and the volume filling fraction is $\sim
8\%$ and decreases with distance due to the flux-limited data. Consequently, filaments
contain the largest fraction of mass in the Universe and they represent the most salient
component of the cosmic web: they form the bridges between all structural features at the
group/cluster scale. 

Our catalogue of filaments is not the first attempt to extract filaments from SDSS data.
Filaments from SDSS have been extracted by \citet{Sousbie:08}, \citet{Jasche:09}, and \citet{Smith:12}. In the
following studies, we plan to compare how various filament finders work and how the
filaments detected using different methods differ.

In our method we have to define the filament scale (radius). In the current study it is
fixed at $0.5~h^{-1}$Mpc. This scale was chosen to find the bridges between galaxy groups
and because it was known that this scale affects the galaxy evolution
\citep[e.g.][]{Tempel:12b,Tempel:13a}. Since filaments are hierarchical by nature
\citep{Aragon-Calvo:10,Smith:12} it is interesting to search for filaments at many scales.
The number density of galaxies in the SDSS does not allow to search for smaller filaments.
We are preparing our filament finder to search for thicker filaments, and the catalogue
and comparison of multi-scale filaments will be our next step. The multi-scale filaments
will allow us to better determine the filament scale that affects galaxy evolution.

Figure~\ref{fig:fraction} shows that galaxies in large clusters are not in filaments. This
is expected since filaments are the bridges between clusters and large clusters are in
intersection of many filaments \citep{Aragon-Calvo:10}. In our following work, we plan to
study how filaments and groups/clusters are connected and how this connection depends on
cluster/filament properties.

The Bisous process can also be applied to other structure elements, as clusters, sheets,
and voids. This is also one the future directions of our work.

%=============================================================================
\section*{Acknowledgments}

We thank our colleagues in the Cosmology Department of Tartu Observatory for useful
discussions that helped to clarify many aspects in the paper. We thank the referee for his/her constructive comments and questions.

We acknowledge the support by the Estonian Science Foundation grants 9428, MJD272, PUT246; the
Estonian Ministry for Education and Science research project SF0060067s08; the Centre of
Excellence of Dark Matter in (Astro)particle Physics and Cosmology; by the Spanish Ministry
of Science and Innovation project AYA2010-22111-C03-02, and by the Generalitat Valenciana
project of excellence Prometeo 2009/064. E.S. acknowledges a visiting professor grant of the
Vicerrectorado de Investigaci\'on y Pol\'itica Cient\'ifica de la Universitat de Val\`encia.
E.T. and E.S. also thank the hospitality of the Observatori Astron\`omic, Universitat de
Val\`encia and Universit\'e Lille~1 where part of this work was performed. R.S. thanks the Universit\'e Lille~1, the GDR~3477 G\'eom\'etrie stochastique, the Tartu Observatory and the Valencia Observatory for the financial and scientifical support. This work was carried out in the High Performance Computing Center of University of Tartu. All the figures
have been made using the gnuplot plotting utility or matplotlib \citep{Hunter:2007}. This research has made use of NASA's Astrophysics Data System Bibliographic Services.

Funding for SDSS-III has been provided by the Alfred P. Sloan Foundation, the Participating Institutions, the National Science Foundation, and the U.S. Department of Energy Office of Science. The SDSS-III web site is http://www.sdss3.org/.

SDSS-III is managed by the Astrophysical Research Consortium for the Participating Institutions of the SDSS-III Collaboration including the University of Arizona, the Brazilian Participation Group, Brookhaven National Laboratory, Carnegie Mellon University, University of Florida, the French Participation Group, the German Participation Group, Harvard University, the Instituto de Astrofisica de Canarias, the Michigan State/Notre Dame/JINA Participation Group, Johns Hopkins University, Lawrence Berkeley National Laboratory, Max Planck Institute for Astrophysics, Max Planck Institute for Extraterrestrial Physics, New Mexico State University, New York University, Ohio State University, Pennsylvania State University, University of Portsmouth, Princeton University, the Spanish Participation Group, University of Tokyo, University of Utah, Vanderbilt University, University of Virginia, University of Washington, and Yale University.

%\bibliographystyle{mn2e}
%\bibliography{mybib}{}

\appendix

\section{Description of the catalogue}
\label{app:1}

The catalogue of filaments consists of three tables. The first table lists the extracted
filaments and the general properties (e.g., length) of the filaments. The second table
gives all the filament points with their properties: every filament from the first table
consists of a point set with a spacing of $\approx 0.5~h^{-1}\mathrm{Mpc}$. The third
table lists the galaxies that we used to generate the filaments: the galaxies are
extracted from \citet{Tempel:12}. The galaxy table lists the basic galaxy properties (for
more properties, please see the Table in \citealt{Tempel:12}) together with the info on
the filament where the galaxy belongs to.

The catalogues are accessible at \url{http://www.aai.ee/~elmo/sdss-filaments/}
with a complete description in the \texttt{readme.txt} file. We give these catalogues as ascii
files as well as a \texttt{fits} table with three extensions, one for each table. We will
also upload the catalogues to the Strasbourg Astronomical Data Center (CDS).

\subsection{Description of the filament catalogue}
\label{sect:a1}

The filament catalogue (see Table~\ref{tab:a1}) contains the following information (the column numbers are given in
square brackets):
\begin{enumerate}
	\renewcommand{\theenumi}{\arabic{enumi}.}
	\item{[1]\,\texttt{id} --} unique identification number for a filament;
	\item{[2]\,\texttt{npts} --} number of points in the filament with a spacing $\sim 0.5~h^{-1}\mathrm{Mpc}$ (the filament consists of these points);
	\item{[3]\,\texttt{len} --} filament length in units of $h^{-1}\mathrm{Mpc}$, measured along the filament from point to point;
	\item{[4--5]\,\texttt{ngal1, ngal2} --} numbers of galaxies in the filament that are closer than 0.5, 1.0~$h^{-1}\mathrm{Mpc}$ to filament axis;
	\item{[6--7]\,\texttt{lum1, lum2} --}  luminosity of the filament: the sum of luminosities of observed galaxies that are closer than 0.5, 1.0~$h^{-1}\mathrm{Mpc}$ to filament axis (in units of $10^{10}h^{-2}L_\odot$);
	\item{[8--10]\,\texttt{xmin, ymin, zmin} --} filament minimum coordinate in $x, y, z$ axis; the coordinates are defined by Eq.~(\ref{eq:coordin});
	\item{[11--13]\,\texttt{xlen, ylen, zlen} --} filament range in $x, y, z$ axis.
\end{enumerate}

\subsection{Description of the filament points table}
\label{sect:a2}

The table of filament points (see Table~\ref{tab:a2}) contains the following information (the column numbers are
given in square brackets):

\begin{enumerate}
	\renewcommand{\theenumi}{\arabic{enumi}.}
	\item{[1]\,\texttt{id} --} filament identification number;
	\item{[2]\,\texttt{idpts} --} unique identification number for a filament point, shared for all filaments; 
	\item{[3]\,\texttt{npts} --} number of filament points in the filament the point belongs to;
	\item{[4]\,\texttt{len} --} length of the filament (in units of $h^{-1}\mathrm{Mpc}$) the point belongs to;
	\item{[5--7]\,\texttt{x, y, z} --} the co-moving coordinates ($x$, $y$, $z$) in units of $h^{-1}\mathrm{Mpc}$ as defined by Eq.~(\ref{eq:coordin});
	\item{[8]\,\texttt{dist} --} distance to the filament point in units of $h^{-1}\mathrm{Mpc}$; 
	\item{[9--11]\,\texttt{dx, dy, dz} --}  orientation of the filament at that point as defined by $\bmath{\omega}_{\mathcal{G}}$, the orientation is given as an unit vector;
	\item{[12]\,\texttt{vmap} --} visit map (level set) value ($\mathcal{L}$);
	\item{[13]\,\texttt{fden} --} weighted visit map value, filament density ($\mathcal{D}$);
	\item{[14]\,\texttt{fori} --} strength of orientation as defined by $\mathcal{D}_\mathcal{G}$.
\end{enumerate}

\subsection{Description of the galaxies table}
\label{sect:a3}

The table of galaxies (see Table~\ref{tab:a3}) contains the following information (the column numbers are given in
square brackets):

\begin{enumerate}
	\renewcommand{\theenumi}{\arabic{enumi}.}
	\item{[1]\,\texttt{id} --} unique identification number for a galaxy, as used in \citet{Tempel:12};
	\item{[2]\,\texttt{nrich} --} richness of the group the galaxy belongs to;
	\item{[3]\,\texttt{redshift} --}  redshift, corrected to the CMB rest frame;
	\item{[4--5]\,\texttt{ra, dec} --}  right ascension and declination (deg);
	\item{[6]\,\texttt{distcor} --}  co-moving distance of the galaxy when the finger-of-god effect is suppressed (as used in filament extraction);
	\item{[7--11]\,\texttt{mag\_$x$} --} Galactic extinction corrected Petrosian magnitude ($x\in ugriz$ filters);
	\item{[12]\,\texttt{lumr} --} absolute luminosity in the $r$-band in units of $10^{10}h^{-2}L_\odot$, where $M_\odot=4.64$ \citep{Blanton:07};
	\item{[13]\,\texttt{w} --} weight factor for the  galaxy (\texttt{w}$\cdot$\texttt{lumr} was used to calculate the luminosity density field);
	\item{[14]\,\texttt{edgedist} --}  co-moving distance of the galaxy from the border of the survey mask;
	\item{[15]\,\texttt{fil\_dist} --} distance from the nearest filament axis (or from filament end point) in units of $h^{-1}\mathrm{Mpc}$;	
	\item{[16]\,\texttt{fil\_id} --} id of the nearest filament;
	\item{[17]\,\texttt{fil\_idpts} --} id of the nearest filament point.
\end{enumerate}

\begin{table*}
 \caption{A sample of filament table. The full table is available online. See Sect.~\ref{sect:a1} for detailed description.}
 \begin{tabular}{@{}ccccccccccccc}
  \hline\hline
  id & npts & len & ngal1 & ngal2 & lum1 & lum2 & xmin & ymin & zmin & xlen & ylen & zlen \\
     &      & $\mathrm{Mpc}/h$ &     &     & $10^{10}L_{\odot}/h^2$ & $10^{10}L_{\odot}/h^2$ &
	 $\mathrm{Mpc}/h$ & $\mathrm{Mpc}/h$ & $\mathrm{Mpc}/h$ & $\mathrm{Mpc}/h$ & $\mathrm{Mpc}/h$ & $\mathrm{Mpc}/h$ \\
  \hline
  1 &  50 &   24.79  & 17 &    22 &    24.57 & 40.07 &-129.64 &  270.94 & -158.41 &  17.97 &  13.90 &  6.47  \\
  2 &  91 &   45.27  & 21 &    25 &    30.94 & 36.40 &-128.90 &  170.87 & -120.65 &  14.51 &  38.65 & 11.60 \\
  3 &  45 &   22.22  & 20 &    28 &     5.85 &  8.98  & 49.34   & 87.33  & -55.79  &  6.29  & 10.46 & 17.61 \\
  4 &  76 &   37.63  & 27 &    29 &    21.99 & 25.18 & -46.78  & 134.59 &   41.37   & 7.24  & 30.41 &  9.61  \\
  5 &  33 &   16.22  & 18 &    25 &     7.33 & 11.01 & -32.24  &  75.18  & -15.86  &  1.79  & 15.40 &  3.31  \\
  \hline
 \end{tabular}
 \label{tab:a1}
\end{table*}

\begin{table*}
 \caption{A sample of filament points table. The full table is available online. See Sect.~\ref{sect:a2} for detailed description.}
 \begin{tabular}{@{}cccccccccccccc}
  \hline\hline
  id & idpts & npts & len & x & y & z & dist & dx & dy & dz & vmap & fden & fori \\
     &    &      & $\mathrm{Mpc}/h$ & $\mathrm{Mpc}/h$ & $\mathrm{Mpc}/h$ & $\mathrm{Mpc}/h$ & $\mathrm{Mpc}/h$ &
	 & & & & & \\
  \hline
  1 & 1 &     50 &   24.79 & -111.66 & 270.94 & -155.93 & 331.96 &  -0.5467 & 0.4396 &  -0.7126 & 0.434 & 0.095 & 0.855 \\
  1 & 2 &     50 &   24.79 & -111.98 & 271.13 & -156.31 & 332.40 &  -0.6738 & 0.5447 &  -0.4991 & 0.530 & 0.139 & 0.836 \\
  1 & 3 &     50 &   24.79 & -112.33 & 271.36 & -156.61 & 332.85 &  -0.7001 & 0.5836 &  -0.4113 & 0.586 & 0.176 & 0.788 \\
  1 & 4 &     50 &   24.79 & -112.68 & 271.61 & -156.87 & 333.29 &  -0.7015 & 0.5523 &  -0.4503 & 0.507 & 0.148 & 0.915 \\
  1 & 5 &     50 &   24.79 & -113.04 & 271.85 & -157.13 & 333.73 &  -0.7198 & 0.5147 &  -0.4657 & 0.451 & 0.131 & 0.974 \\
  \hline
 \end{tabular}
 \label{tab:a2}
\end{table*}

\begin{table*}
 \caption{A sample of galaxy table. The full table is available online. See Sect.~\ref{sect:a3} for detailed description.}
 \begin{tabular}{@{}ccccccccccccc}
  \hline\hline
  id & nrich & redshift & ra & dec & distcor &  mag\_r  & lumr &
  w & edgedist & fil\_dist & fil\_id & fil\_idpts \\
    &  &  & deg & deg & $\mathrm{Mpc}/h$ & mag & $10^{10}L_{\odot}/h^2$ & & $\mathrm{Mpc}/h$ & $\mathrm{Mpc}/h$ & & \\
   
  \hline
  16 &   2 &    0.1044 & 251.16 & 28.22 & 308.57 &  16.73 &  1.61 & 1.81 & 10.79 &  2.59 &  6778 &   177435 \\   
  17 &   2 &    0.1062 & 251.17 & 28.13 & 309.10 &  17.40 &  0.91 & 1.81 & 11.05 &  2.54 &  6778 &   177435 \\   
  18 &   3 &    0.1324 & 251.34 & 28.46 & 387.90 &  17.68 &  1.13 & 2.55 & 11.37 &  0.12 &  218 &    10252 \\    
  19 &   3 &    0.1337 & 251.35 & 28.48 & 388.02 &  17.56 &  1.25 & 2.55 & 11.29 &  0.06 &  218 &    10253 \\    
  22 &   3 &    0.1331 & 251.33 & 28.49 & 387.96 &  17.66 &  1.11 & 2.55 & 11.32 &  0.07 &  218 &    10253 \\    
  \hline
 \end{tabular}
 \label{tab:a3}
\end{table*}

\label{lastpage}

\end{document}